\documentclass[twocolumn,showpacs,preprintnumbers,amsmath,amssymb,prb]{revtex4}
\usepackage{graphicx}
\usepackage{dcolumn}
\usepackage{bm}
\usepackage[usenames,dvipsnames]{color}

\newcommand {\SPSSe} {Sn$_2$P$_2$(Se$_x$S$_{1-x}$)$_6$}
\newcommand {\SPS}    {Sn$_2$P$_2$S$_6$}
\newcommand {\SPSe}    {Sn$_2$P$_2$Se$_6$}

\begin{document}


\title{ Nonequilibrium effects near the Lifshitz point in uniaxial ferroelectrics with strongly anharmonic local potential}

\title{Observation of Kibble-Zurek behavior near the Lifshitz point in ferroelectrics with incommensurate phase}

\author{K.Z.~Rushchanskii}
\email{k.rushchanskii@fz-juelich.de}
\affiliation {Peter Gr\"unberg Institut, Forschungszentrum J\"ulich and JARA, D-52425 J\"ulich, Germany}

\author{A.~Molnar}
\affiliation{Institute for Solid State Physics and Chemistry, Uzhgorod National University, 54 Voloshyn St., 88000 Uzhgorod, Ukraine}

\author{R.~Bilanych}
\affiliation{Institute for Solid State Physics and Chemistry, Uzhgorod National University, 54 Voloshyn St., 88000 Uzhgorod, Ukraine}

\author{R.~Yevych}
\affiliation{Institute for Solid State Physics and Chemistry, Uzhgorod National University, 54 Voloshyn St., 88000 Uzhgorod, Ukraine}

\author{A.~Kohutych}
\affiliation{Institute for Solid State Physics and Chemistry, Uzhgorod National University, 54 Voloshyn St., 88000 Uzhgorod, Ukraine}

\author{Yu.M.~Vysochanskii}
\affiliation{Institute for Solid State Physics and Chemistry, Uzhgorod National University, 54 Voloshyn St., 88000 Uzhgorod, Ukraine}

\author{V.~Samulionis}
\affiliation{Physics Faculty, Vilnius University, Sauletekio al. 9/3, 10222 Vilnius, Lithuania}

\author{J.~Banys}
\affiliation{Physics Faculty, Vilnius University, Sauletekio al. 9/3, 10222 Vilnius, Lithuania}

\date{\today}

\begin{abstract}
We have investigated non-equilibrium properties of proper uniaxial  Sn$_2$P$_2$(Se$_x$S$_{1-x}$)$_6$ ferroelectrics with the Type II incommensurate phase above Lifshitz point  $x_{\rm LP} \sim 0.28$. We measured dielectric susceptibility with cooling and heating rate ranging 0.002-0.1~K/min, and high-resolution ultrasound experiments and hypersound Brillouin scattering. For samples with $x \geqslant 0.28$ clear anomalies were observed at incommensurate second order  transition ($T_i$) and at first order lock-in transition ($T_c$) in the regime of very slow cooling rate, whereas  the intermediate IC phase is not observed when the rate is faster then 0.1~K/min.  In general, increasing the cooling rate leads to smearing the anomaly at $T_c$. We explain this effect in terms of Kibble-Zurek model for non-equilibrium second order phase transitions. In the ferroelectrics with strongly nonlinear local potential cooling rate defines concentration of domain walls and their size:  domain width decreases when cooling rate increases. At certain conditions the size of domain is comparable to the incommensurate phase modulation period, which lies in micrometer scale in the vicinity of Lifshitz point and leads to pinning of the modulation period by domain wall.
\end{abstract}

\maketitle

\section{INTRODUCTION}
Multiferroic properties and functionality of nano-sized objects are related to inhomogeneous space distribution of the order parameters close to the surface and to competitive interactions in the bulk of materials.
This could lead to exotic domain structures \cite{bubble_domains} and/or to modulation waves in periodic polarization arrangement\cite{ref1, ref2} below phase transition (PT) temperature (i.e., in polar phase).
Strong nonlinearity of the local potential could determine peculiar shape of domain walls and their temperature evolution in ferroelectric phase. \cite{ref6, ref7, ref8}
When domain dimension starts to be comparable with modulation wave length in the incommensurate phase, we could expect their interference leading to new interesting phenomena. In materials with the Lifshitz point \cite{ref3, ref4, ref5} (LP, see below) in their compositional (or pressure) -- temperature phase diagrams the modulation period could be changed continuously. This make them a good candidates to study such interference.
Subsequently, interesting phenomena could appear by transformation of domain structure into incommensurate (IC) modulation across lock-in transition near the LP.

Such possibilities could be investigated in unique case of \SPSSe\ mixed crystals.\cite{ref2} From one side, sulphide  \SPS\ (SPS) is proper uniaxial ferroelectric with the second order PT at $T_0 \sim 337$~K from monoclinic $P2_1/c$ paraelectric phase to $Pc$ ferroelectric one. Strongly nonlinear local potential for the polarization fluctuations determines mixed displacive--order/disorder nature of this PT.\cite{ref10, ref11, ref12} Corresponding isostructural selenide  \SPSe\  has IC phase between second order transition at $T_i \sim 221$~K and first order lock-in transition at $T_c \sim 193$~K. In pure \SPSe\ the IC phase has modulation period about 14 elementary cells. \cite{ref13,ref14} In mixed \SPSSe\ compounds the IC phase appears at $x > x_{\rm LP} \sim 0.28$, where the $x_{\rm LP}$ denotes the LP concentration on $T-x$ diagram.\cite{ref5,ref15,ref16} When concentration of selenium $x$ decreases then both modulation wave number $q_i$ and temperature range of the IC phase $T_i-T_c$  continuously go to zero as $q_i \sim (x-x_{\rm LP})^{0.5}$ and $T_i-T_c \sim (x- x_{\rm LP})^2$ correspondingly, as is expected in the mean field theory of the LP with one component order parameter and one direction of modulation.\cite{ref4} The LP demonstrates new universality class for critical anomalies of thermodynamic properties. \cite{ref3} The critical behavior could be modified for the case of uniaxial ferroelectrics \cite{ref17} (uniaxial Lifshitz point, ULP) and for the coincidence with position of tricritical point (uniaxial tricritical Lifshitz point, ULTP).\cite{ref18a, ref18b, ref19} For composition $x = 0.28$  high precision heat diffusion experiments\cite{ref20} revealed, that critical exponents in paraelectric and ferroelectric phases as well as ratio of amplitudes are well described by the LP theory, where long-range dipole interactions are not accounted. This observation is related to significant screening of dipole interactions in \SPSe\ with relatively small fundamental bang gap, and, therefore, significantly high concentration of free charge carriers.

The thermodynamic anomalies in the IC phase above the lock-in PT at $T_c$ for \SPSe\ were described taking into account the high nonlinearity of the local potential.\cite{ref7,ref8} In Ref.~\onlinecite{ref2} theoretical and experimental studies of the temperature dependence of spontaneous polarization inhomogeneity revealed typical width of the domains  near several tens of micrometers. The temperature variation of the domain structure also reveals some metastability in SPS. \cite{ref21}

Inhomogeneous interaction of the spontaneous polarization with deformations  was accounted in phenomenological theory of ferroelectrics with Type~II IC phase, which includes the Lifshitz-like invariant in the Landau thermodynamic potential.\cite{ref22a,ref22b,ref23}
Such interaction could be observed as linear coupling of the soft optic and acoustic branches and have been studied by neutron scattering for \SPS\ and \SPSe\ crystals.\cite{ref14,ref24} This interaction was also observed in \SPSSe\ mixed crystals near the LP, where clear evidence of the softening of longitudinal and transverse acoustic branches  have been established in Brillouin scattering and ultrasound experiments. \cite{ref25}

The IC phase and LP diagram could be described in axial next-nearest neighbor interaction (ANNNI) model,\cite{ref26} which accounts for the ratio between strength of nearest and next nearest interactions. Substitution of sulfur by selenium increases the chemical bonds covalence\cite{ref27a,ref27b} and, therefore, rearranges the intercell interactions. In the phonon spectra this replacement leads to decrease of the LO-TO splitting for the lowest energy polar soft-mode and to strengthen of linear interaction of the soft optic and acoustic phonons, which are polarized in the monoclinic symmetry plane.\cite{ref28}

   Similar mesoscopic scales appear near the LP: (i) for the modulation period of IC phase, which is determined by interatomic interactions, and (ii) for the domain width in ferroelectric phase, that additionally depends on macroscopic conditions. Therefore, interesting nonequilibrium effects could be expected in nearest vicinity of the LP.

 Such non-equilibrium effects in the second order phase transitions are described by Kibble-Zurek (KZ) model,\cite{ref29,ref30} which relates the concentration of the symmetry breaking topological defects with the the cooling rate across the PT. As it was shown for molibdates with two component order parameter, the domain walls represent the vector topological defects.\cite{ref31}  Experimentally obtained  domain walls concentration as function of cooling rate is in agreement with KZ model.

In our case of uniaxial ferroelectrics the scalar topological defects are presented as domain walls in the ferroelectric phase. For \SPSSe\ crystals the anomaly of dielectric susceptibility near the lock-in transition is determined by both the domain walls concentration in ferroelectric phase and wave number modulation inside the IC phase. The later could be varied by concentration $x$ when approaching to the LP. Previously, it was found, that near the LP the IC phase temperature interval could be broadened by laser radiation.\cite{ref32}  In this papers we study non-equilibrium effects at different cooling rates in the crystals with different $x$ values.  We will show, that both incommensurate second order (at $T_i$) and lock-in first order (at $T_c$) transitions are still observed for $x = 0.28$ sample when the lowest cooling rate of 0.002~K/min is fulfilled. This means that the LP concentration coordinate is expected at smaller selenium content, probably near $x = 0.26$. The lock-in transition anomaly disappears when cooling rate rises up to 0.1~K/min. Therefore, we point out that determination of the LP coordinates is affected by conditions of the experimental investigations.

The reminder of this paper is organized as follows: in Sec.~II we describe experimental details of our study, Sec.~III contains results of dielectric measurements, ultrasound and Brillouin scattering experiments, in Sec.~IV we discuss these results in terms of KZ model. We conclude in Sec.~V.

\section{EXPERIMENTAL DETAILS}
We performed accurate study of dielectric susceptibility, ultrasound velocity and Brillouin scattering. Special attention was paid for control of the temperature. The dielectric susceptibility was investigated utilizing digital Goodwill LCR-815 high-end  LCR meter at frequency $10$~kHz. Variation of temperature was in the range $0.1$ to $0.002$~K/min.

The measurements of the ultrasound velocity were performed using computer controlled pulse-echo equipment.\cite{ref34} The precision of relative velocity measurements was better than $10^{-4}$. The temperature stabilization in ultrasound experiments was better than 0.02~K. The sample was carefully polished to have precisely parallel faces (100). Silicone oil was used as the acoustic coupling medium for longitudinal waves. The measurements were carried out at 10~MHz frequency using piezoelectric LiNbO$_3$ transducers.

The Brillouin scattering was investigated in back-scattering geometry using a He-Ne laser and a pressure-scanned three-pass Fabry-P\'{e}rot interferometer\cite{ref25} with sharpness of 35 and free spectral range of 2.51~cm$^{-1}$. The samples were placed in a UTREX cryostat in which the temperature was stabilized with an accuracy of 0.3~K.

Single crystals of the \SPSSe\  were grown by the vapor transport (VT) and Bridgeman (BR) technologies. The content of sulfur and selenium in obtained samples is in good agreement with their nominal values.\cite{ref15, ref33} All investigated samples were prepared as plates with $5\times5\times3$~mm$^3$ dimensions. Silver paste electrodes were attached to the largest (001) face, which is nearly normal to the direction of spontaneous polarization.

\section{RESULTS}

\begin{figure*}
\includegraphics[width=8.6cm]{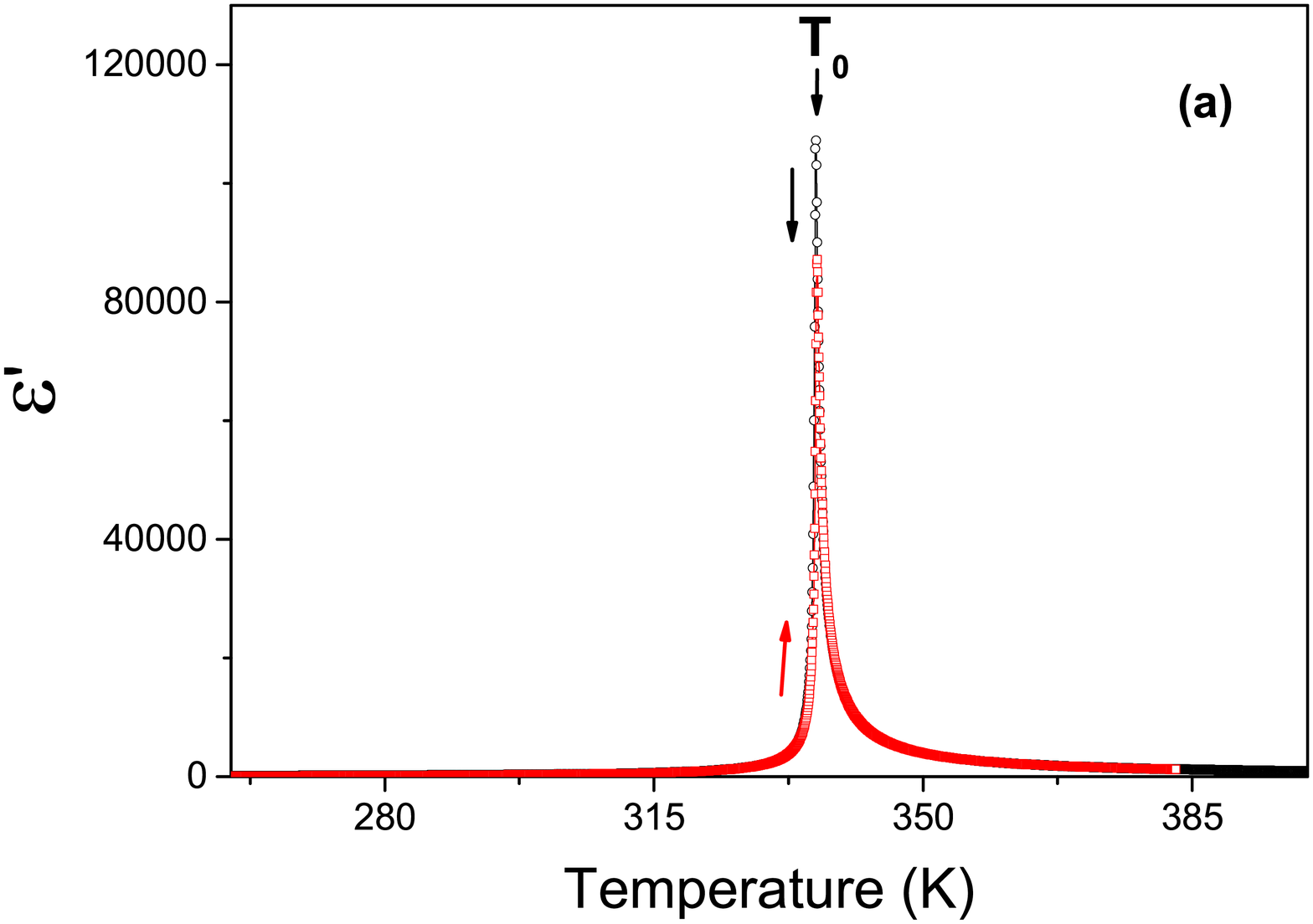}%
 \includegraphics[width=8.6cm]{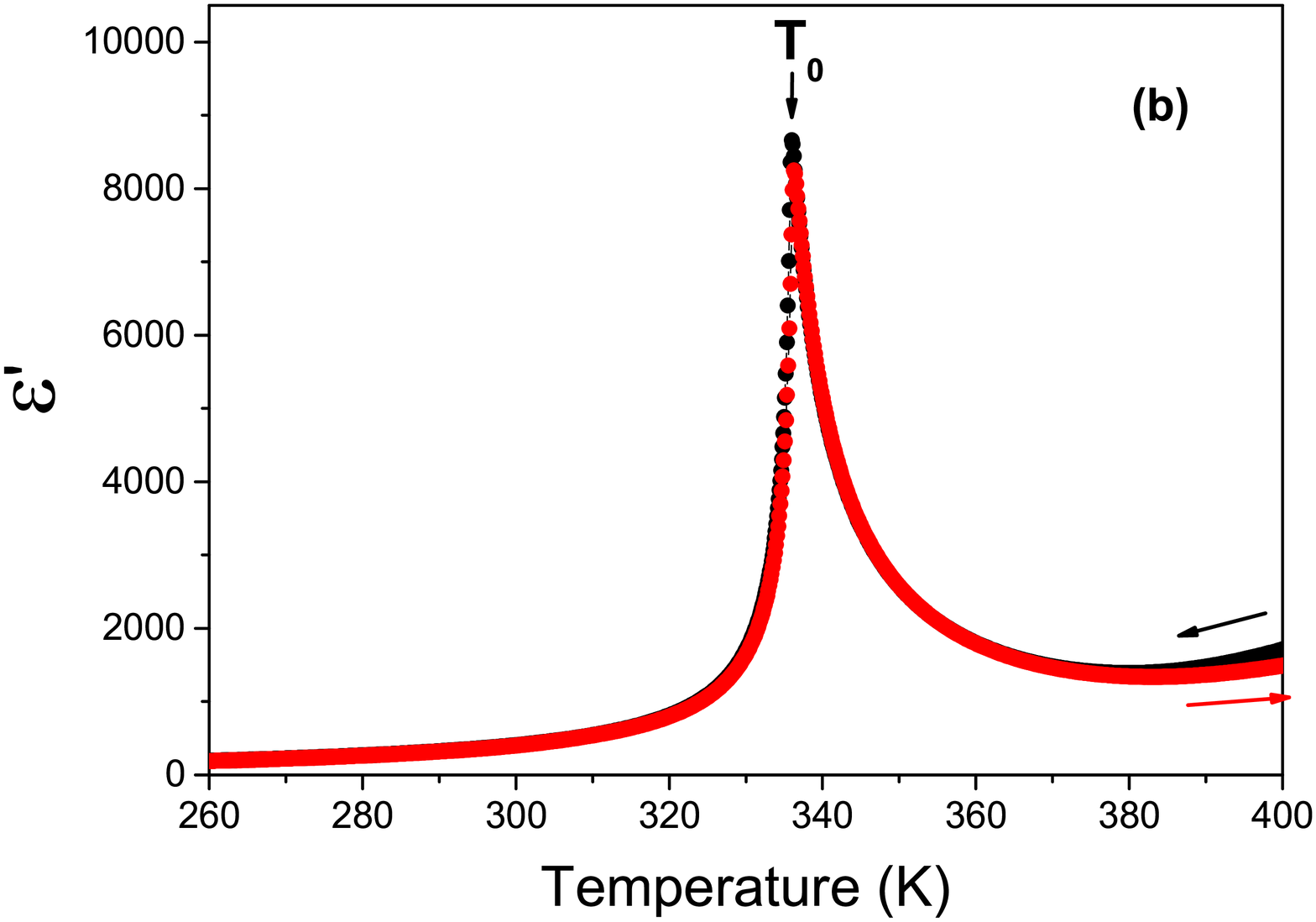} \\
  \includegraphics[width=8.6cm]{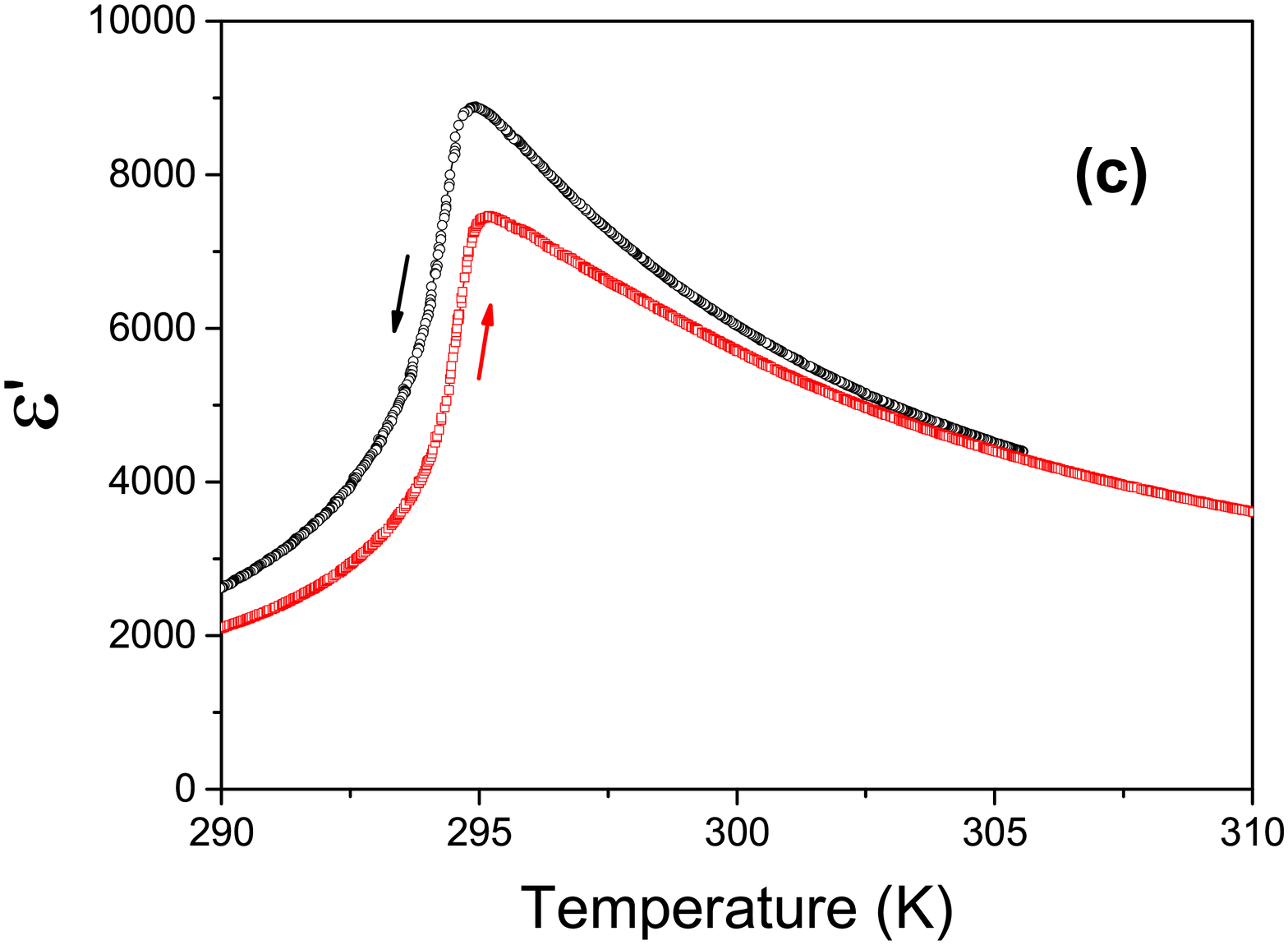}%
\includegraphics[width=8.6cm]{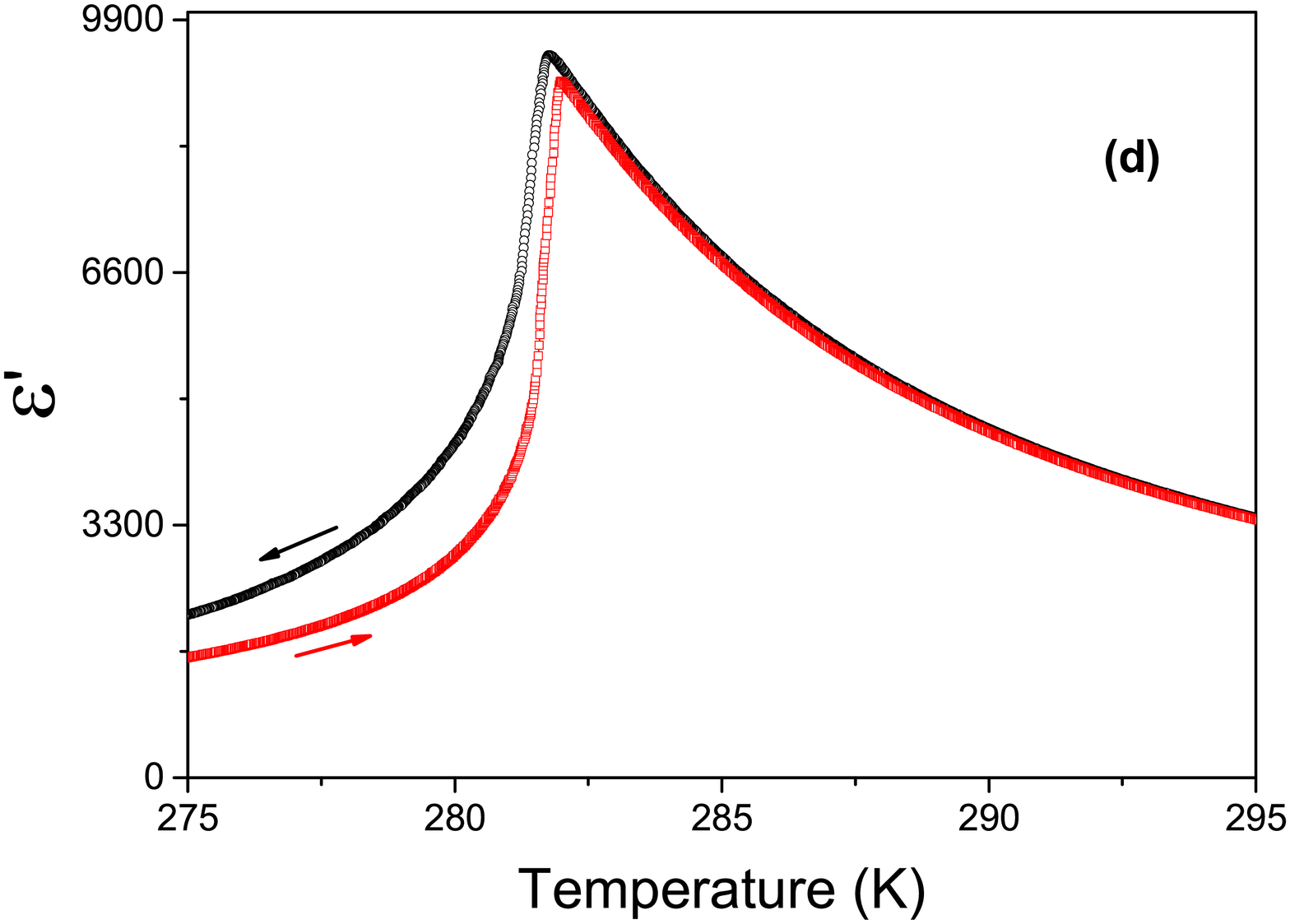} \\
 \includegraphics[width=8.6cm]{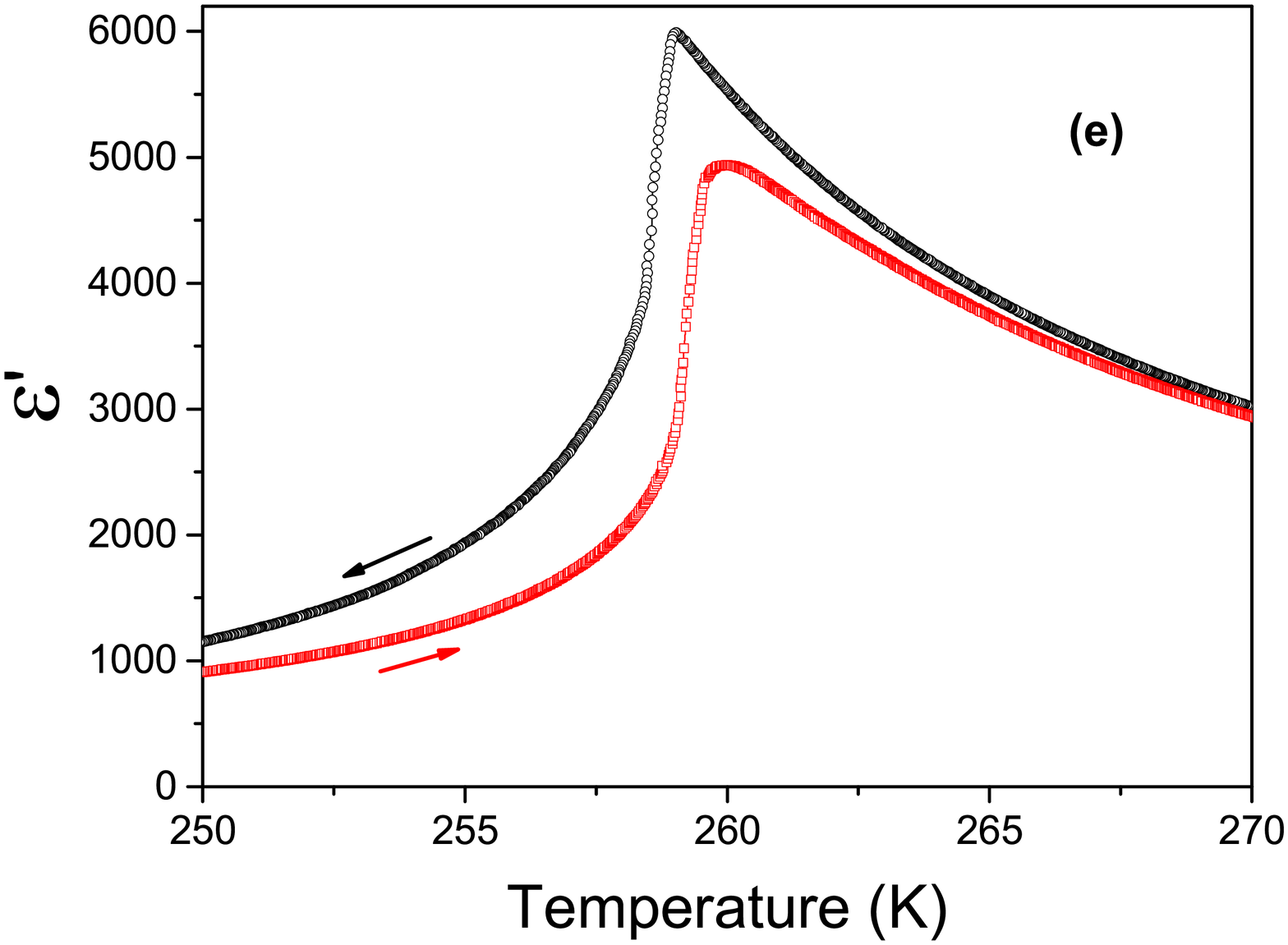}%
 \includegraphics[width=8.6cm]{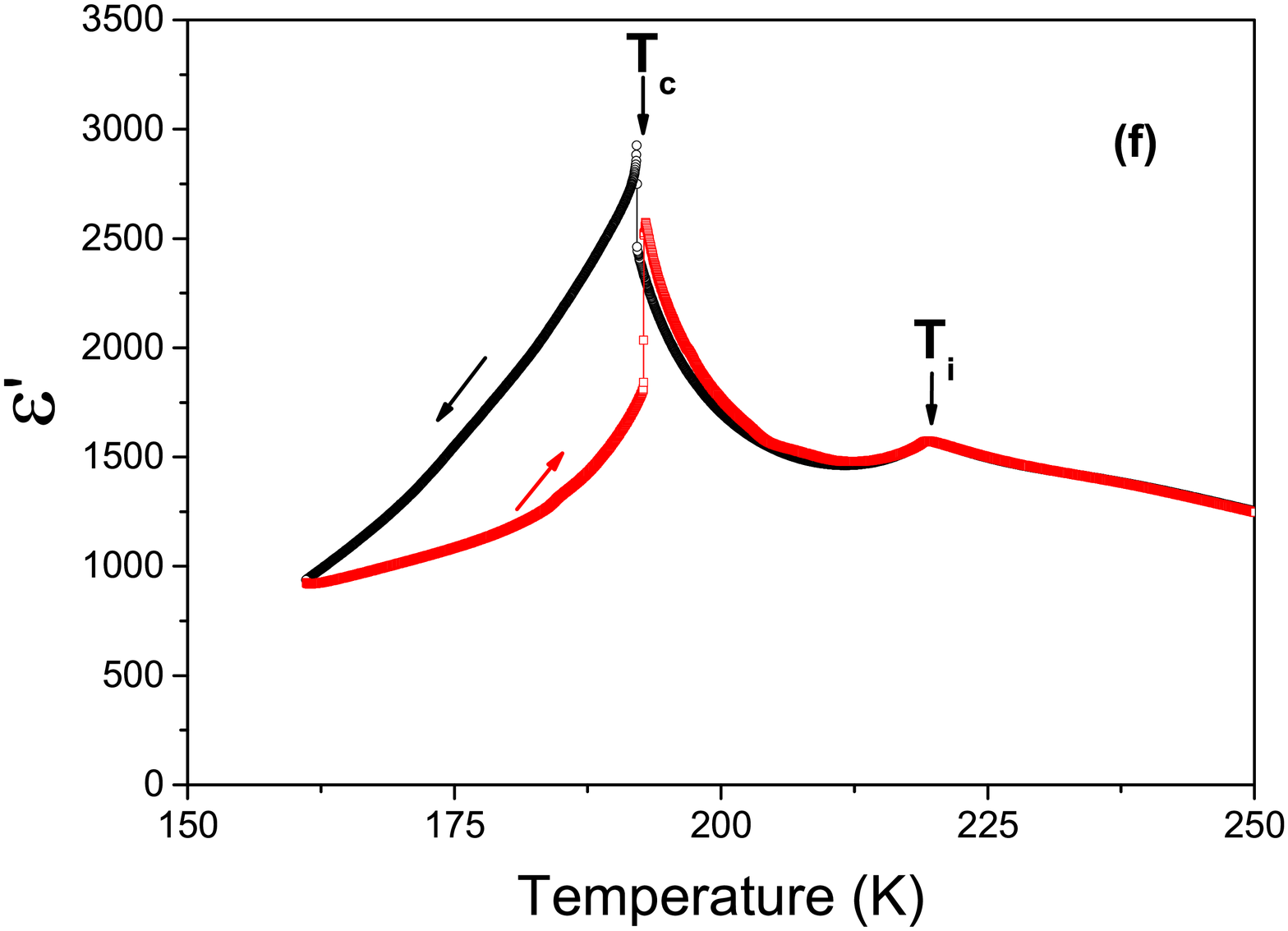} \\
\begin{minipage}[c]{0.5\textwidth}
    \includegraphics[width=8.6cm]{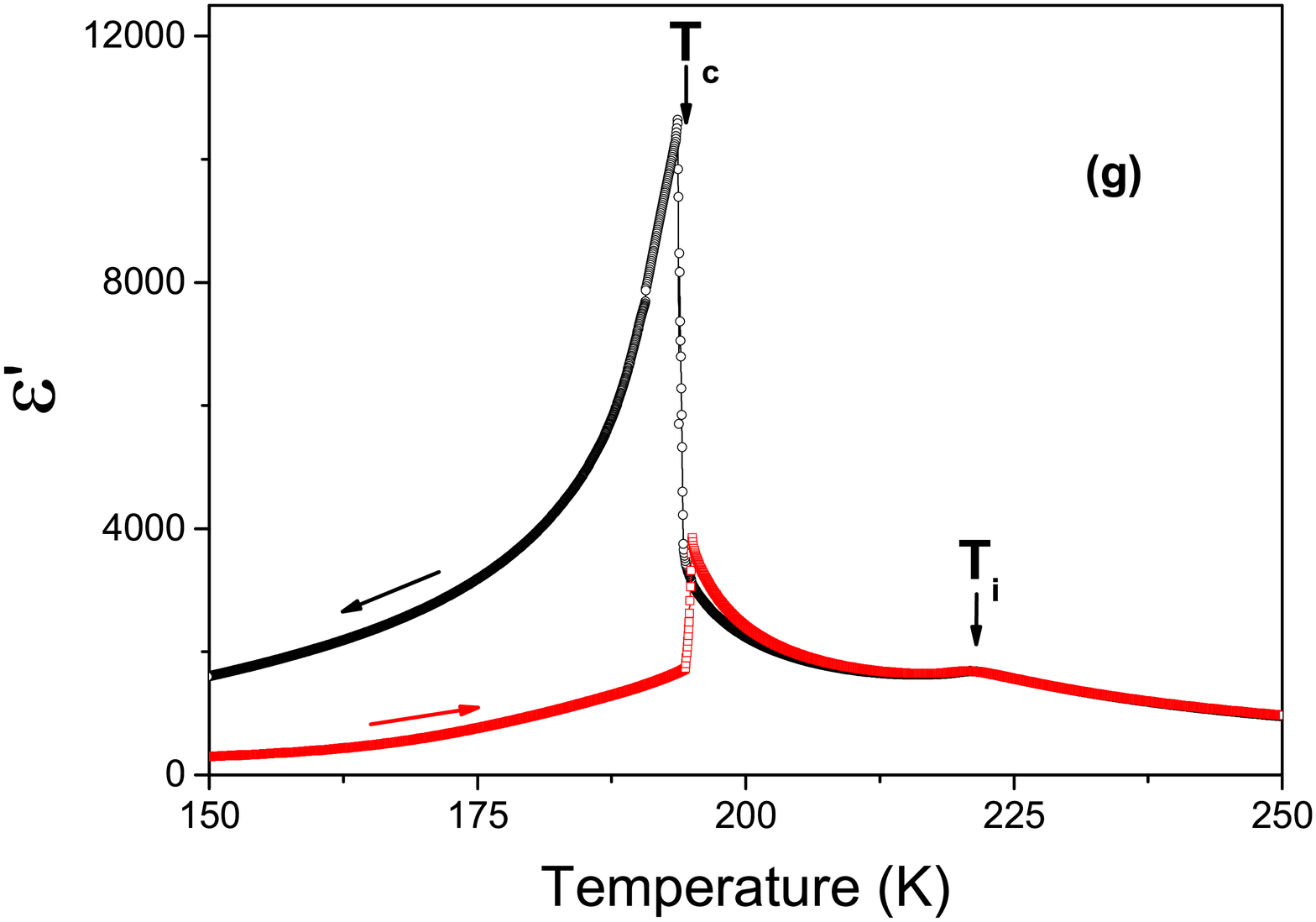}
  \end{minipage}\hfill
  \begin{minipage}[c]{0.48\textwidth}
    \caption{The temperature dependencies of dielectric susceptibility for Sn$_2$P$_2$(Se$_x$S$_{1-x}$)$_6$ crystals measured at 0.005 K/min cooling and heating rate.
(a)  $x = 0$, BR sample;
(b)  $x = 0$, VT sample;
(c)  $x = 0.22$;
(d)  $x = 0.28$;
(e) $x = 0.4$;
(f)  $x = 1$, VT smple;
(g) $x = 1$, BR sample.
    } \label{Fig1}
  \end{minipage}
\end{figure*}
The temperature dependence of dielectric susceptibility for \SPSSe\ crystals, that were grown by VT method, are shown at Figure~1. For the samples with compositions $x = 0$, 0.22 and 0.28 the dielectric susceptibility at $T_0(x)$ was reached its maximal values about 8000. For mixed crystal with $x = 0.4$ the maximum of susceptibility decreases: they reaches only near 6000. For \SPSe\  two dielectric anomalies are seen: with maximal susceptibility about 3000 at $T_c$ and about 1500 at $T_i$.

For  \SPS\ and \SPSe\ crystals grown by BR technology  the dielectric susceptibility in paraelectric phase obeys Curie-Weiss temperature dependence with Curie constant $C = (0.6 - 0.7) \cdot 10^5$~K. The susceptibility reaches very high values (above $10^5$) at ferroelectric second order transition ($T_0$) in sulfide compound and near $3 \cdot 10^4$ at first order lock-in transition $T_c$ in selenide crystal (see Figure~1). The difference in observed maxima of dielectric susceptibility in VT and BR samples is related to different contribution of domain walls to dielectric susceptibility in samples with different conductivity: VT samples are more conductive.\cite{ref35} Also, observed temperature hysteresis is related to contribution of domain walls. For slow variation of temperature (0.005 K/min) the hysteresis in the position of   maxima  by cooling and heating is about 0.2~K for the BR sample and it is little bigger (about 0.25~K) for the VT \SPS\ sample.

The peculiarity is also clearly reflected in temperature dependence of dielectric susceptibility  by cooling and heating in ferroelectric phase of both BR and VT \SPSe\ samples (Figure~1).  The temperature hysteresis of the maxima of dielectric susceptibility at first order lock-in transition ($T_c$) is about 1 and 2~K for VT and BR \SPSe\   samples, correspondingly.

The temperature hysteresis inside incommensurate phase of BR and VT \SPSe\ samples doesn't demonstrate so clear difference as in the case of domain walls dielectric contribution.
Temperature hysteresis is not observed in paraelectric to incommensurate transition ($T_i$) for both VT and BR \SPSe\ samples.

\begin{figure*}
\includegraphics[width=8.6cm]{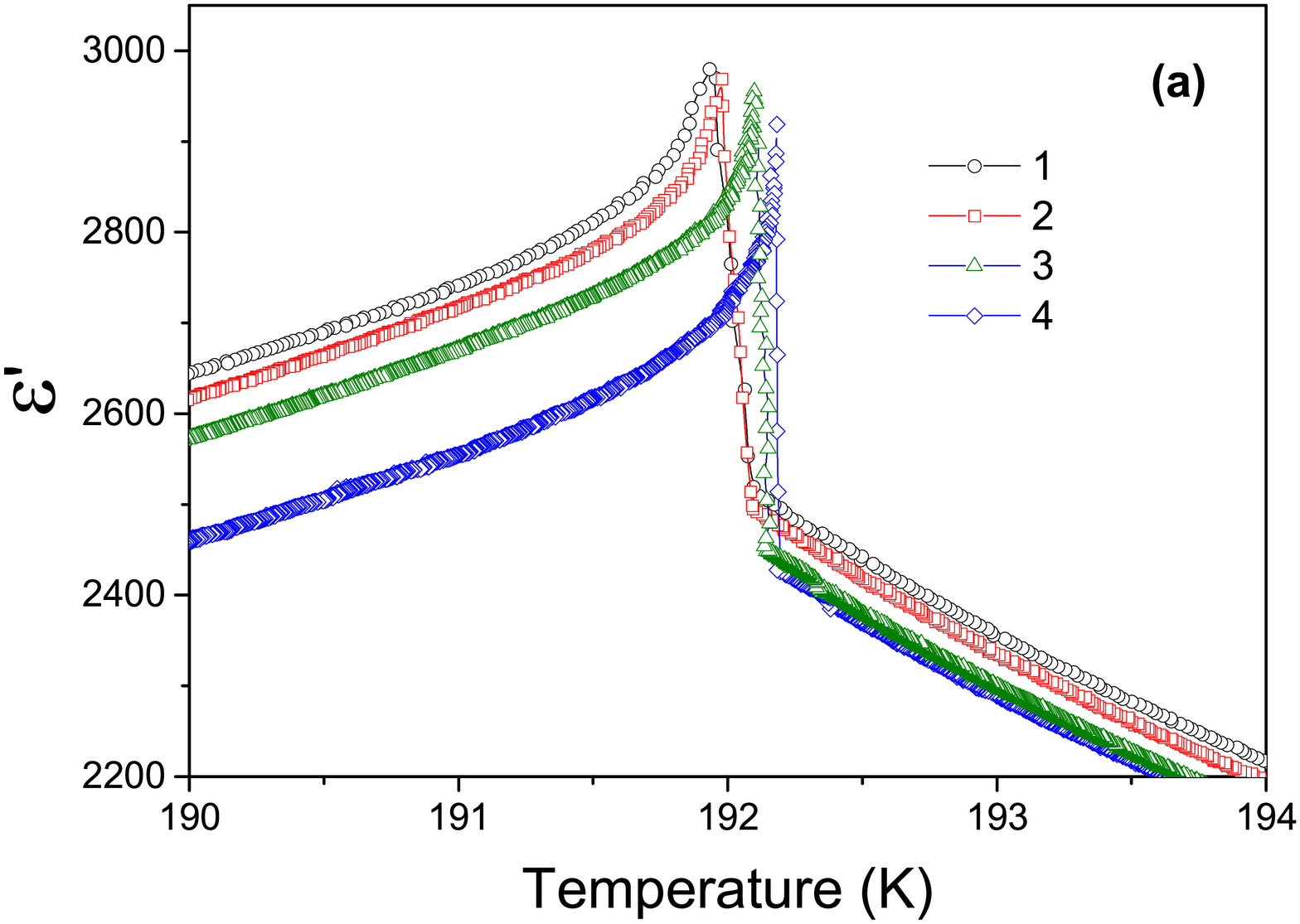}%
\includegraphics[width=8.6cm]{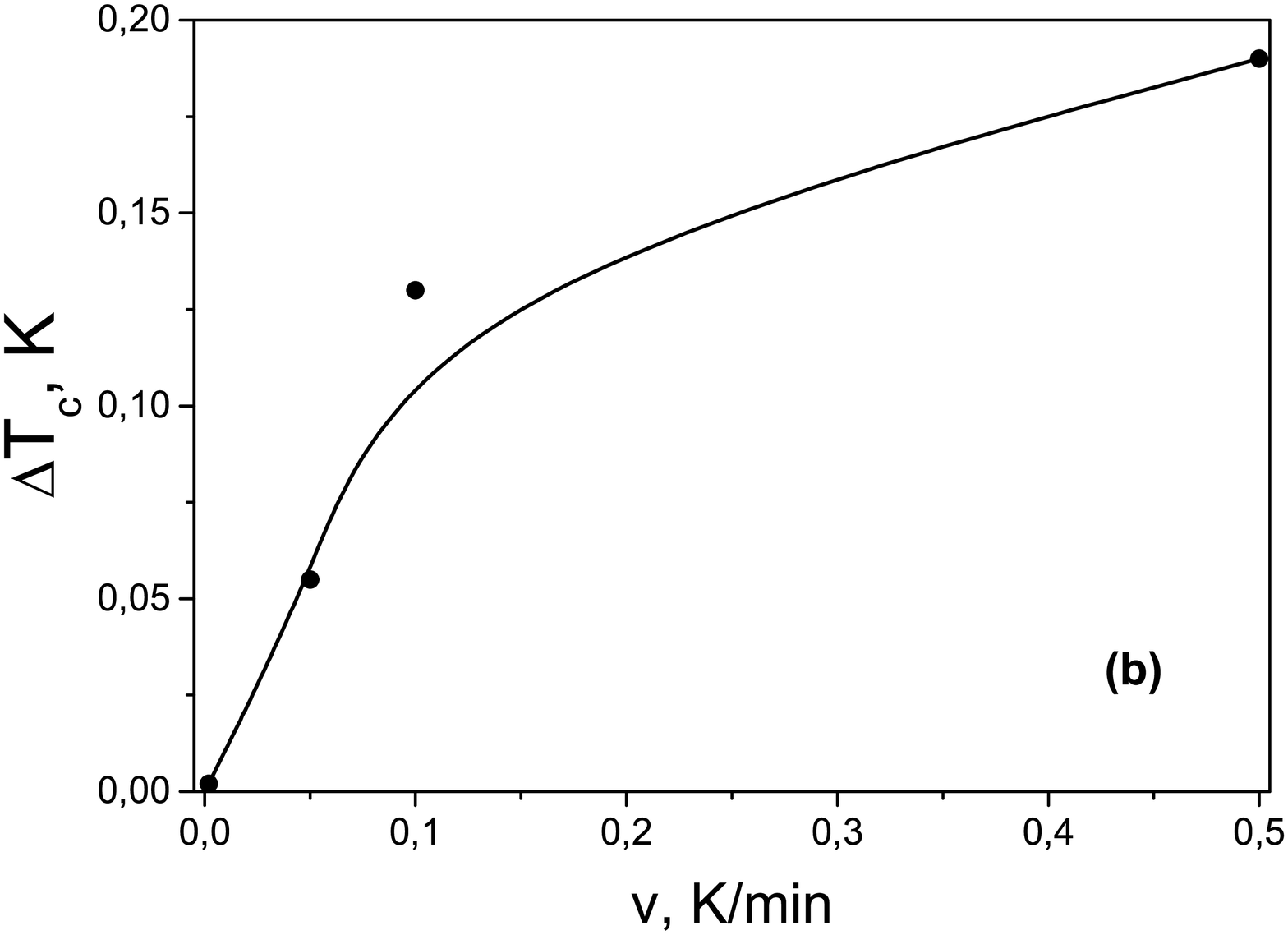} \\
\begin{minipage}[c]{0.5\textwidth}
    \includegraphics[width=8.6cm]{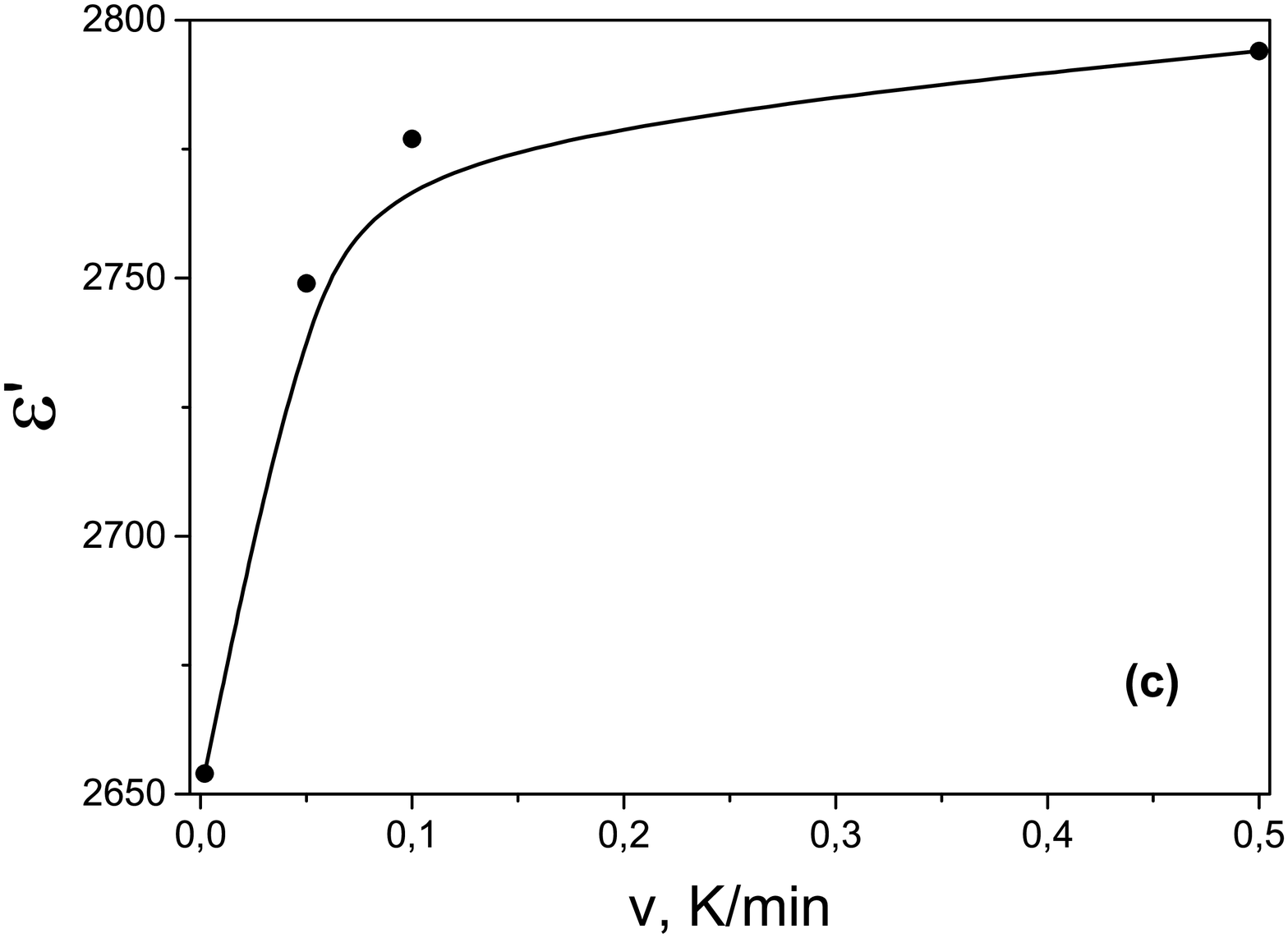} \\
  \end{minipage}\hfill
  \begin{minipage}[c]{0.48\textwidth}
    \caption{ (a) The temperature dependence of dielectric susceptibility for \SPSe VT sample at the following cooling rates: (1) 0.5 K/min, (2) 0.1 K/min, (3) 0.05 K/min, and (4) 0.005 K/min;
 The cooling rate dependences of the (b) lock-in transition temperature width and (c) dielectric susceptibility at $T_c - 0.5$~K.} \label{Fig2}
  \end{minipage}
\end{figure*}
In any case, the dielectric behavior near lock-in transition in \SPSe\ crystals is strongly dependent on regime of measurements. The temperature width $\Delta$T$_c$ of the lock-in transition increases from 0.002~K at the cooling rate 0.005 K/min till 0.13~K at the rate 0.1~K/min, and further to 0.19~K at the rate 0.5~K/min (see Figure~2). The dielectric susceptibility in ferroelectric phase, for example at 0.5~K below $T_c$ (see Figure~2b) also  growth rapidly when cooling rate increases from 0.005 to 0.1~K/min. For higher rates the dielectric susceptibility is almost constant.

These data demonstrate, that contribution of domain walls to the dielectric susceptibility growth  at $T_i$ due to increase of their concentration at higher cooling rates. Some increase of dielectric susceptibility in the low temperature range of the incommensurate phase (see Figure~2a, together with widening of the lock-in transition temperature interval) could be also related to higher concentration of the modulation wave defects. These defects are areas with new modulation period (so named nucleations or stripples, see Ref.~\onlinecite{ref36a, ref36b}), which appeared by faster cooling.

Naturally, the following question appears:  What is the influence  of the cooling rate on dielectric properties across the incommensurate phase with small temperature interval and with weak first order lock-in transition in the nearest vicinity of LP?
In this vicinity the modulation wave period is large and could be comparable with the size of domains in the ferroelectric phase just below $T_c$. According to Kibble-Zurek model,\cite{ref29,ref30} if the cooling across the second order phase transition is fast, the concentration of symmetry breaking topological defects (which are the domain walls in the case of ferroelectrics) will be higher. The model  determines the size of domains as the function of the cooling rate, which allow the possibility that the size of domains will be near or equal to the  IC modulation wave length. Therefore, IC modulation wave could be pinned to the domain structure. These pinning will be observed experimentally as smeared anomalies of dielectric susceptibility in the vicinity of the LP. It is known, that  near the LP the phase transitions lines $T_i(x)$ and $T_c(x)$ (which are the borders of the incommensurate phase) must tangentially coincide with the $T_0(x)$ line of direct transition from paraelectric phase into ferroelectric one.\cite{ref4} Therefore, the correct phase diagram in the vicinity of the LP can be obtained only at  experimental conditions which are close to equilibrium. In real experiments such conditions could be satisfied by sufficiently slow variation of temperature, especially in cooling regime. For this aim we have investigated acoustic properties of \SPSSe\ with high resolution in temperature changes and  temperature behavior of dielectric susceptibility at the cooling (heating) rates slowed down to 0.002~K/min.

For the beginning we generalize available data about the temperature-concentration $T-x$ phase diagram for the \SPSSe\ mixed crystals. Our new data concerning the temperature positions of dielectric anomalies are compared with known data obtained in various experiments. Previously, the phase transitions in \SPSSe\  crystals were investigated by optical absorption,\cite{ref33}  X-ray diffraction, \cite{ref13,ref15} neutron scattering,\cite{ref14} heat capacity, \cite{ref37} heat diffusion,\cite{ref20} Brillouin scattering,\cite{ref25,ref28} ultrasound,\cite{ref25,ref34} and dielectric susceptibility\cite{ref38} measurements.  The  $T-x$ diagram, that includes large set of available experimental data, is presented in Figure~3. This diagram indicates at $x_{\rm LP} \sim 0.28$ the presence  of a triple point (namely, the LP), at which the line $T_0(x)$ of second order phase transitions from paraelectric to  ferroelectric phases is continuously split into line $T_i(x)$ of the second order transitions and line $T_c(x)$ of the first order transitions. Between this lines the IC phase is observed. The wave number of incommensurate modulation $q_i$, which appears along $T_i(x)$ line, decreases when the LP is approached. The concentration dependencies\cite{ref15} of the IC phase temperature width $T_i -T_c$ and modulation wave number $q_i$ (see Figure~4) could be interpolated by relations $T_i - T_c \sim (x - x_{\rm LP})^2$ and $q_i \sim (x - x_{\rm LP})^{0.5}$, in agreement with predictions made by the mean field theory.\cite{ref4} In the case of one component order parameter and one direction of the IC modulation  it is expected an inflection of the border line of paraelectric phase  at the LP.\cite{ref3,ref4} Therefore, one could expect different curvatures for the $T_0(x)$ and $T_i(x)$ segments. Indeed, from the polynomial fitting for all set of experimental data on $T_0$ and $T_i$ values in \SPSSe\ crystals such inflection is clearly seen (see Figure~5), and the inflection point is placed near $x = 0.22$.

\begin{figure}
\includegraphics[width=8.6cm]{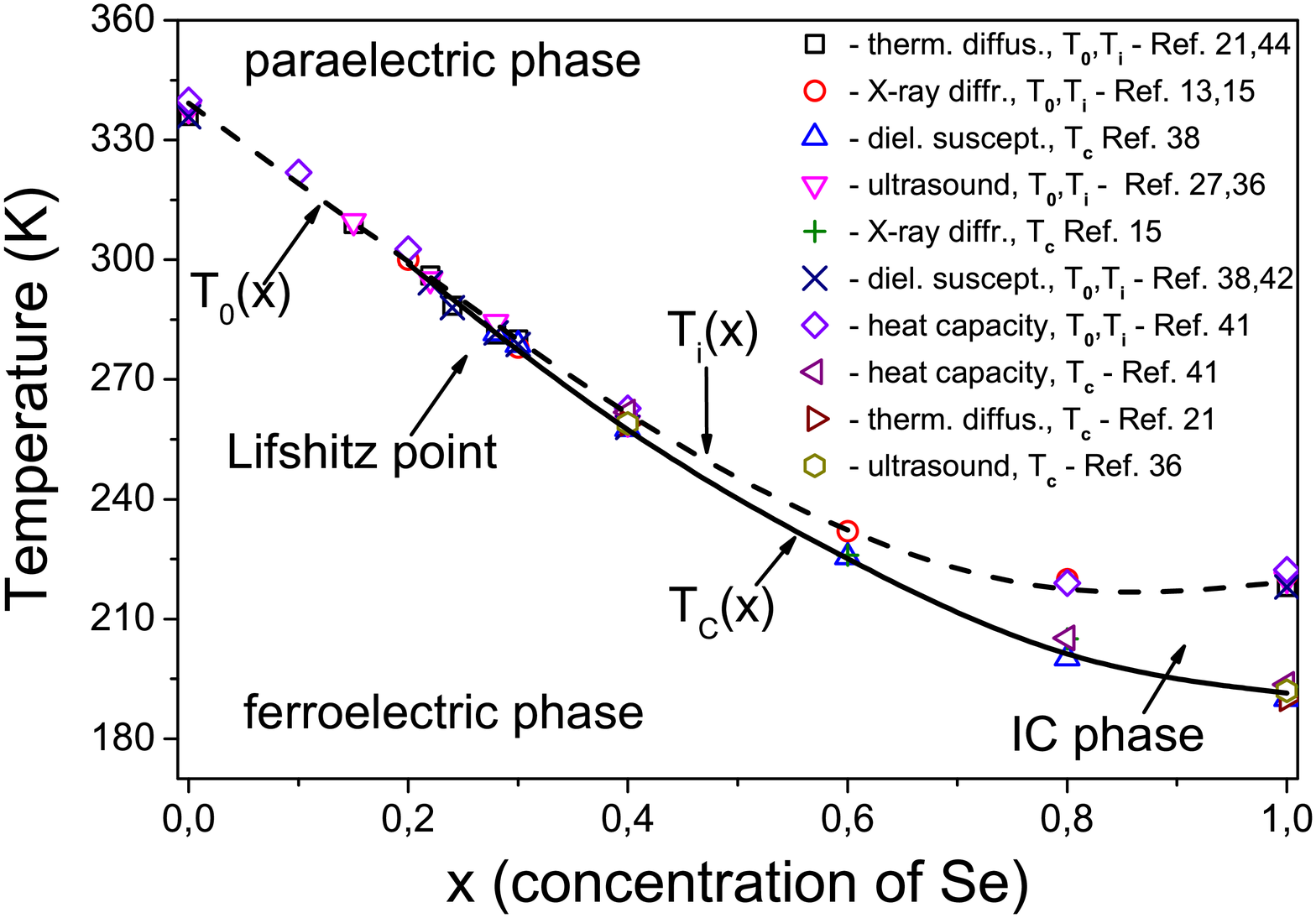} \\
\caption{Phase diagram of \SPSSe\ ferroelectrics.  The second order paraelectric--ferroelectric $T_0(x)$ (at $x < x_{\rm LP}$) and paraelectric--IC $T_i(x)$ (at $x > x_{\rm LP}$) transitions are shown by dashed line. By solid line the first order lock-in transitions between IC and ferroelectric phases are presented. Different style points present experimental data (Ref.\onlinecite{ref13,ref15,ref20,ref25,ref35,ref34,ref38,ref41,ref37}) for the transitions temperatures $T_0$, $T_i$ and $T_c$.}
\label{Fig3}
\end{figure}

\begin{figure}
\includegraphics[width=8.6cm]{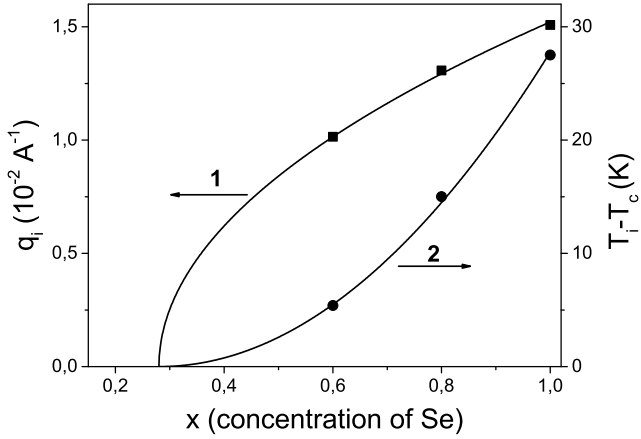} \\
\caption{The concentration dependencies of the modulation wave number along $T_i(x)$ line (1) and temperature width $T_i-T_c$ of IC phase (2) for \SPSSe\ crystals. Points -- experimental data from Ref.~\onlinecite{ref15}. Lines -- approximations by relations  $q_i \sim (x-x_{\rm LP})^{0.5}$   and $T_i - T_c \sim (x-x_{\rm LP})^{2}$.}
\label{Fig4}
\end{figure}

\begin{figure}
 \includegraphics[width=8.6cm]{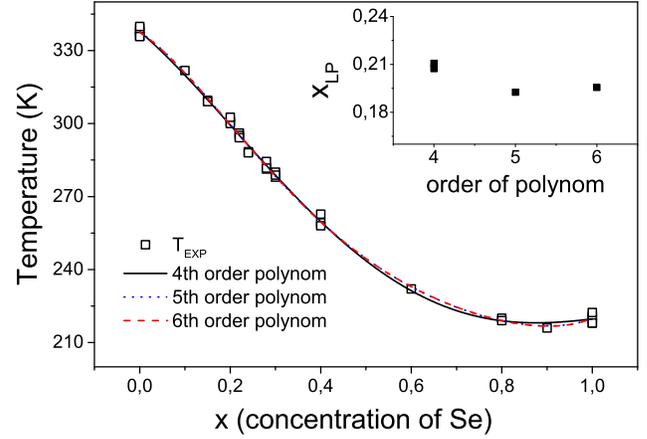} \\
\caption{Fitting of $T_0(x)$ (at $ x < x_{\rm LP}$) and $T_i(x)$ (at $ x > x_{\rm LP}$) phase transitions lines on \SPSSe\ diagram by polynoms of 4th, 5th and 6th order. By squares the experimental data for $T_0$ and $T_i$ (from Figure~4) are shown. Inset: the inflexion point concentration (that could be related to the LP concentration) in dependence of polinom order is presented.}
\label{Fig5}
\end{figure}

The position of the LP could be also evaluated from evolution of the temperature dependence of phonon spectra at different concentration $x$. Previously, neutron scattering study\cite{ref14} revealed linear interaction of low energy soft optic and acoustic phonon branches in \SPSe\  crystal. This interaction was phenomenologically  described as Lifshitz-like invariant in the thermodynamic potential function for proper uniaxial ferroelectrics with Type II IC phase.\cite{ref14,ref16,ref22a,ref22b,ref23}
Such linear interaction near the LP results in a softening of acoustic phonons, what have been observed for the \SPSSe\  crystals in Brillouin scattering and ultrasound investigations. \cite{ref25} Naturally, one could expect, that the acoustic branches softening in some part of reciprocal space near the Brillouin zone center will be reflected in the phonon contribution to the thermal transport. Indeed, according to the thermal diffusion data, \cite{ref20}  the thermal conductivity in paraelectric phase of \SPSSe\  crystals lowers when sulfur is substituted  by selenium  (see Figure~6). But such lowering occurs only till concentration of selenium reaches $x = 0.22$, what is smaller than expected selenium content in mixed crystal with the LP composition $x_{\rm LP} \sim  0.28$. Such peculiarity could be explained by the fact, that short-length phonons are dominated in the heat transport, because they have relatively big concentration (i.e., high density of states) and high group velocity. Therefore, softening of acoustic phonons in the narrow vicinity of the LP doesn�t influences the thermal transport.

\begin{figure}
 \includegraphics[width=8.6cm]{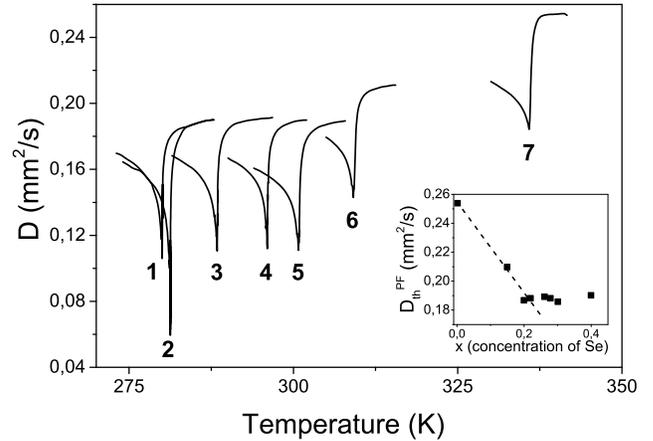} \\
\caption{Temperature anomalies of heat diffusion coefficient in \SPSSe crystals with (1) $x = 0.30$ , (2) $x=0.28$,  (3) $x=0.26$, (4) $x=0.22$, (5) $x=0.20$, (6) $x=0.15$,  (7) $x=0$, according to data from Ref.~\onlinecite{ref20}. Inset illustrates concentration dependence of heat diffusion coefficient in paraelectric phase.}
\label{Fig6}
\end{figure}

\begin{figure}
 \includegraphics[width=8.6cm]{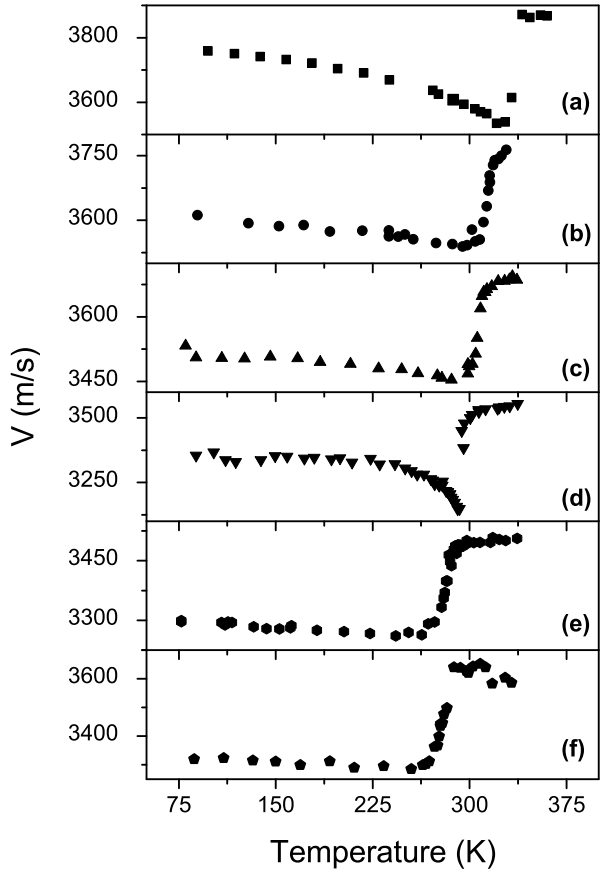} \\
\caption{The temperature dependence of longitudinal hypersound velocity obtained by Brillouin scattering in Z(X X)-Z geometry for \SPSSe\, mixed crystals with (a) $x = 0$, (b) $x=0.10$, (c) $x=0.15$, (d) $x=0.22$, (e) $x=0.28$, (f) $x=0.30$}
\label{Fig7}
\end{figure}

\begin{figure}
 \includegraphics[width=8.6cm]{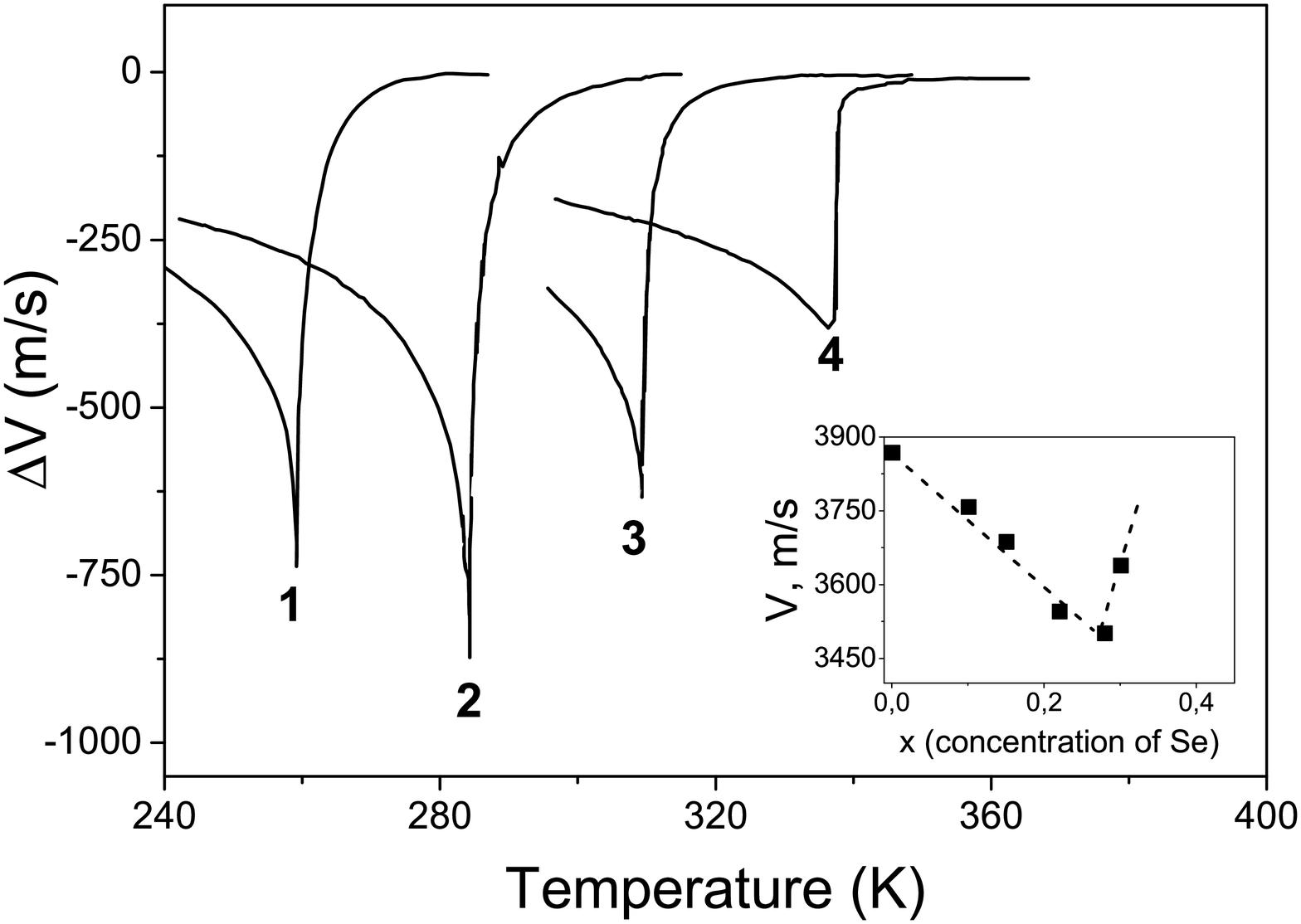} \\
\caption{Temperature variation of longitudinal ultrasound velocity at phase transition \SPSSe crystals with (1) $x = 0.4$, (2) $x=0.28$, (3) $x= 0.15$, (4) $x=0$. Inset illustrates concentration dependence of hypersound velocity in paraelectric phase according data from Figure~7. }
\label{Fig8}
\end{figure}

Results of our Brillouin scattering investigations are presented at Figure~7 as the temperature dependence of  the hypersound velocity. For \SPSSe\ mixed crystals the temperature dependence of longitudinal hypersound in [001] direction is similar to recently reported\cite{ref25,ref28} compositions  $x = 0$ and $x=0.28$. Here, we present results for extended set of compositions to determine concentration dependence of hypersound velocity in paraelectric phase. It appears, that such velocity found minimal value near expected composition $x_{\rm LP}  = 0.28$.

The deepest minimum in the temperature dependence of ultrasound speed is observed  at the composition $x = 0.28$, which is related to the LP position (see Figure~8). Similar deepest minimum in temperature dependence of heat diffusion was observed exactly at $x = 0.28$ of \SPSSe\  mixed crystals.\cite{ref20} For $x = 0.3$ composition the anomaly of ultrasound velocity becomes less deep due to temperature hysteresis of the IC phase and first order lock-in transition.

 \begin{figure}
 \includegraphics[width=8.6cm]{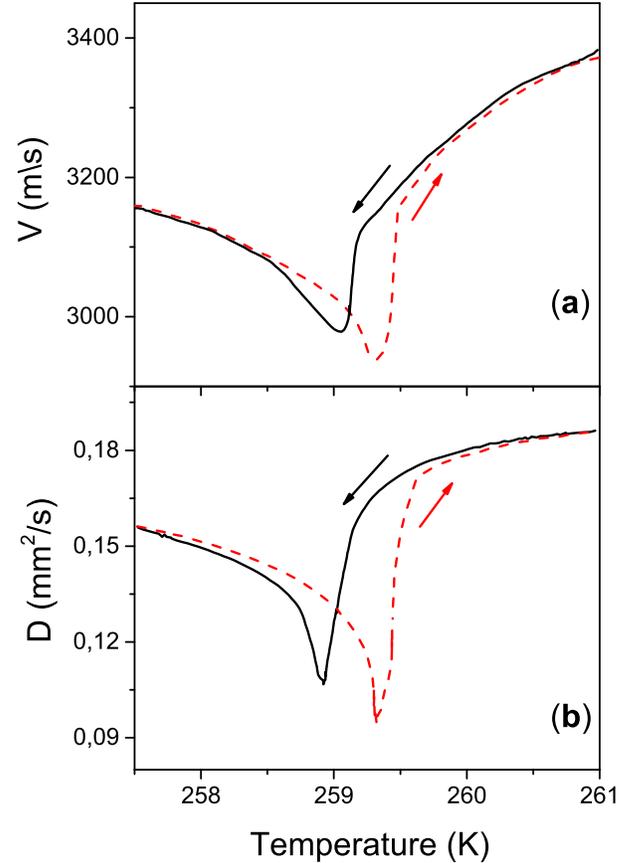} \\
\caption{Temperature dependence of  (a) longitudinal ultrasound velocity, and  (b) heat diffusion coefficient\cite{ref20} by cooling and heating across the phase transitions in \SPSSe\ crystals with $x = 0.4$.}
\label{Fig9}
\end{figure}

The IC phase broadening  is clearly observed\cite{ref20} in the heat diffusion anomaly for sample with $x = 0.4$ composition (see Figure~9b). Our ultrasound data for the same composition $x = 0.4$ (Figure~9a) show, that the temperature anomalies of sound velocity in cooling and heating regimes are very similar to the heat diffusion anomalies in similar regimes. Here, the temperature hysteresis about 0.3~K (see Figure~9b) appears, which is an evidence of strong enough first order lock-in transition at $T_c$. This hysteresis is comparable with temperature interval of IC phase $T_i - T_c$ for this composition.

In general, the acoustic data (which provides information about phase velocity of the ultrasound or hypersound waves) and heat diffusion data (which characterize group velocity of the shortes-waves phonons), together with general shape of $T-x$ diagram for \SPSSe\  mixed crystals (which is based on wide set of different experiments), give an evidence, that the LP should be placed at selenium concentration smaller than $x = 0.28$, somewhere in interval $0.22 < x < 0.28$.

Further, we will discuss the high precision dielectric data, that could help to localize the LP and check possible non equilibrium effects in their vicinity. Clear Curie-Weiss behavior is observed in reciprocal dielectric susceptibility in both paraelectric and ferroelectric phases of \SPS\ crystal with second order PT at $T_0$ (see Figure~10). For \SPSe\ crystal the Curie-Weiss behavior is seen only in the paraelectric phase. Almost symmetric maximum of $1/\epsilon(T)$ dependence is related to continuous PT ($T_i \sim 221$~K) from paraelectric into IC phase. Clear discontinuity in dielectric susceptibility occurs at first order lock-in transition ($T_c \sim 193$~K). From these data follows, that in selenide compound the IC phase is ranged in temperature interval $T_i - T_c \sim 28$~K. For mixed crystals the non-monotonic behavior  in $1/\epsilon(T)$ dependence is also observed, which could be used to define the borders of IC phase with temperature width $T_i - T_c$ near 0.7~K for $x = 0.4$ sample, and near 0.25~K for $x = 0.28$ sample. In case of $x = 0.22$ mixed crystal the dependence $1/\epsilon(T)$ already shows Curie-Weiss behavior in paraelectric and ferroelectic phases with some overshot in ferroelectric phase just below $T_0$ (see Figure~10). So, at low cooling rate the IC phase is clearly observed for the $x = 0.4$ mixed crystal and also is observed in very small temperature interval for $x = 0.28$ sample. But, at faster cooling the jump on $1/\epsilon(T)$ dependence smears for this sample ($x = 0.28$)  (see Figure~11) and dielectric anomaly becomes similar to one found  in the  case of $x = 0.22$ concentration.

\section{DISCUSSION OF EXPERIMENTAL DATA}
For the beginning let�s analyze the phase diagram near the LP. As it was mentioned above, for one component order parameter the mean field approximation predicts, \cite{ref4} that temperature interval of the IC phase has parabolic concentration dependence $T_i - T_c \sim (x - x_{\rm LP})^2$. With account of fluctuations effects the phase diagram is described by critical index $\Phi = 0.625$ in relation $T_i - T_c \sim (x - x_{\rm LP})^{\frac{1}{\Phi}}$.\cite{ref27a, ref27b} It is seen, that in the mean field approximation $\Phi = 0.5$. Also, the concentration behavior of modulation wave number $q_i$ along $T_i(x)$ transitions line follows the relation $q_i \sim (x - x_{\rm LP})^{ \beta_q}$ with index $\beta_q = 0.5$ in the mean field approach. \cite{ref4} The critical behavior near the LP in uniaxial ferroelectrics could be modified by long-range dipole interactions.\cite{ref17} Also, possible proximity to the tricritical LP point  is reflected in new universality class in uniaxial ferroelectrics. \cite{ref18a, ref18b, ref19}

\begin{figure*}
\includegraphics[width=8.6cm]{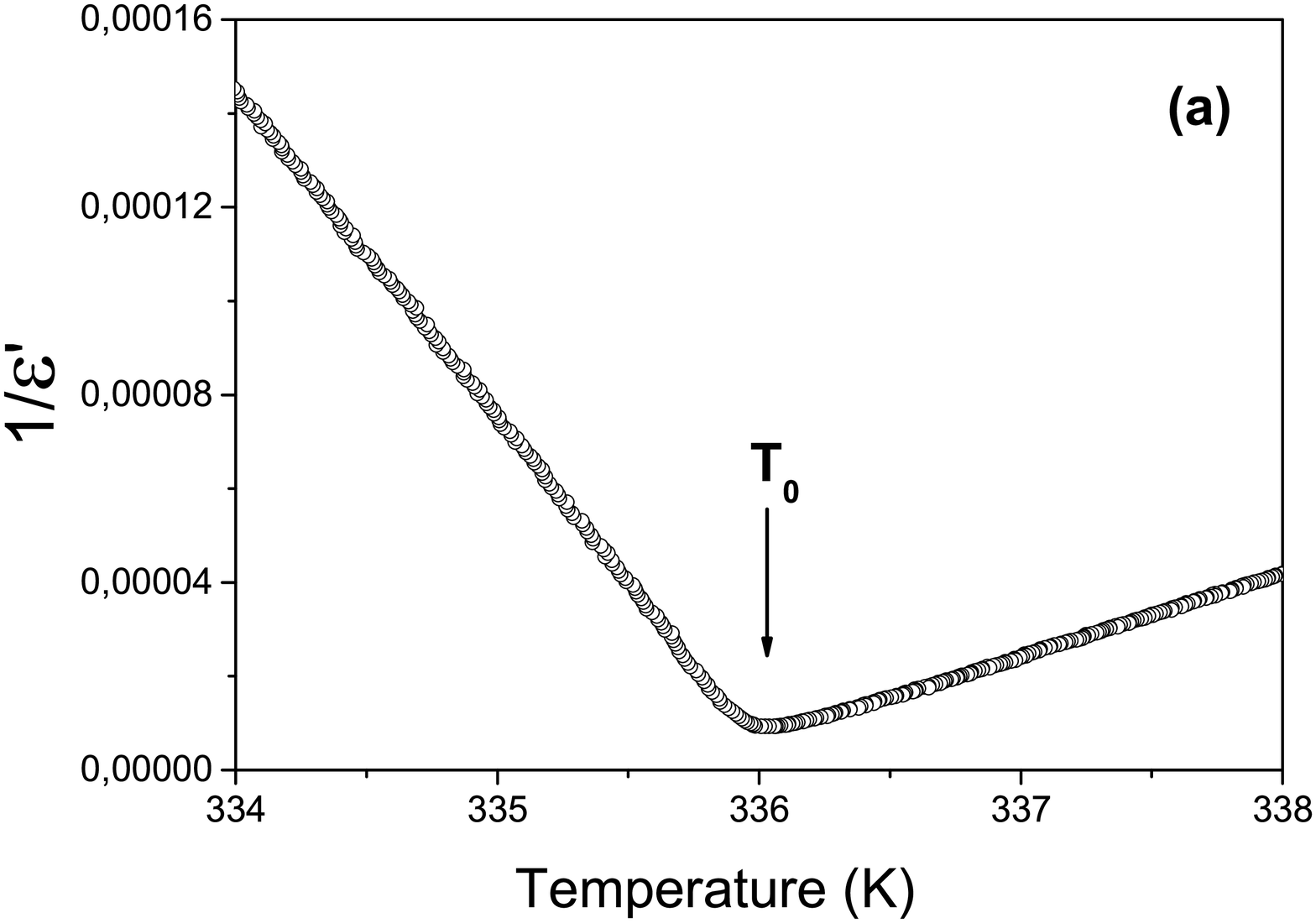}%
\includegraphics[width=8.6cm]{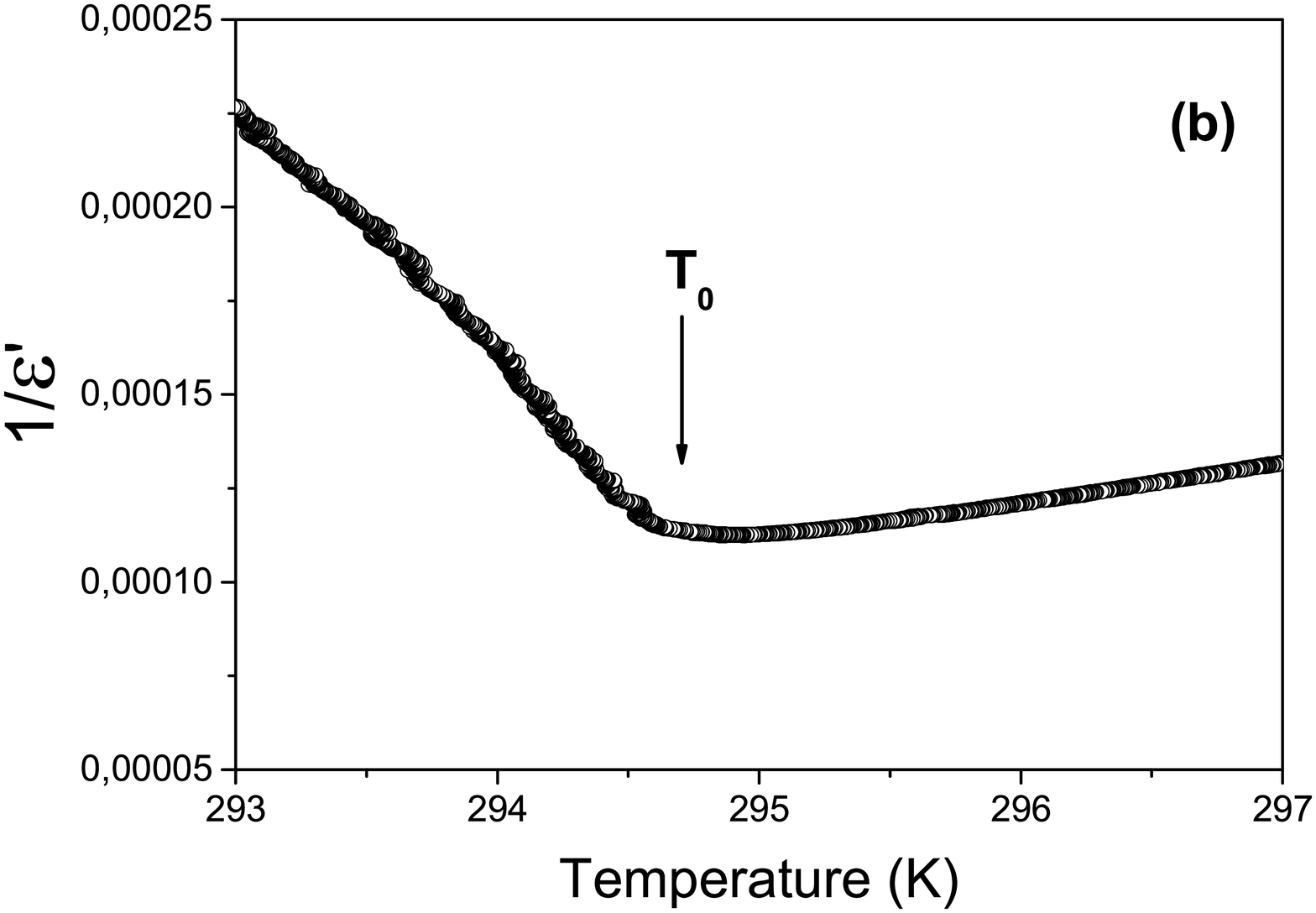} \\
\includegraphics[width=8.6cm]{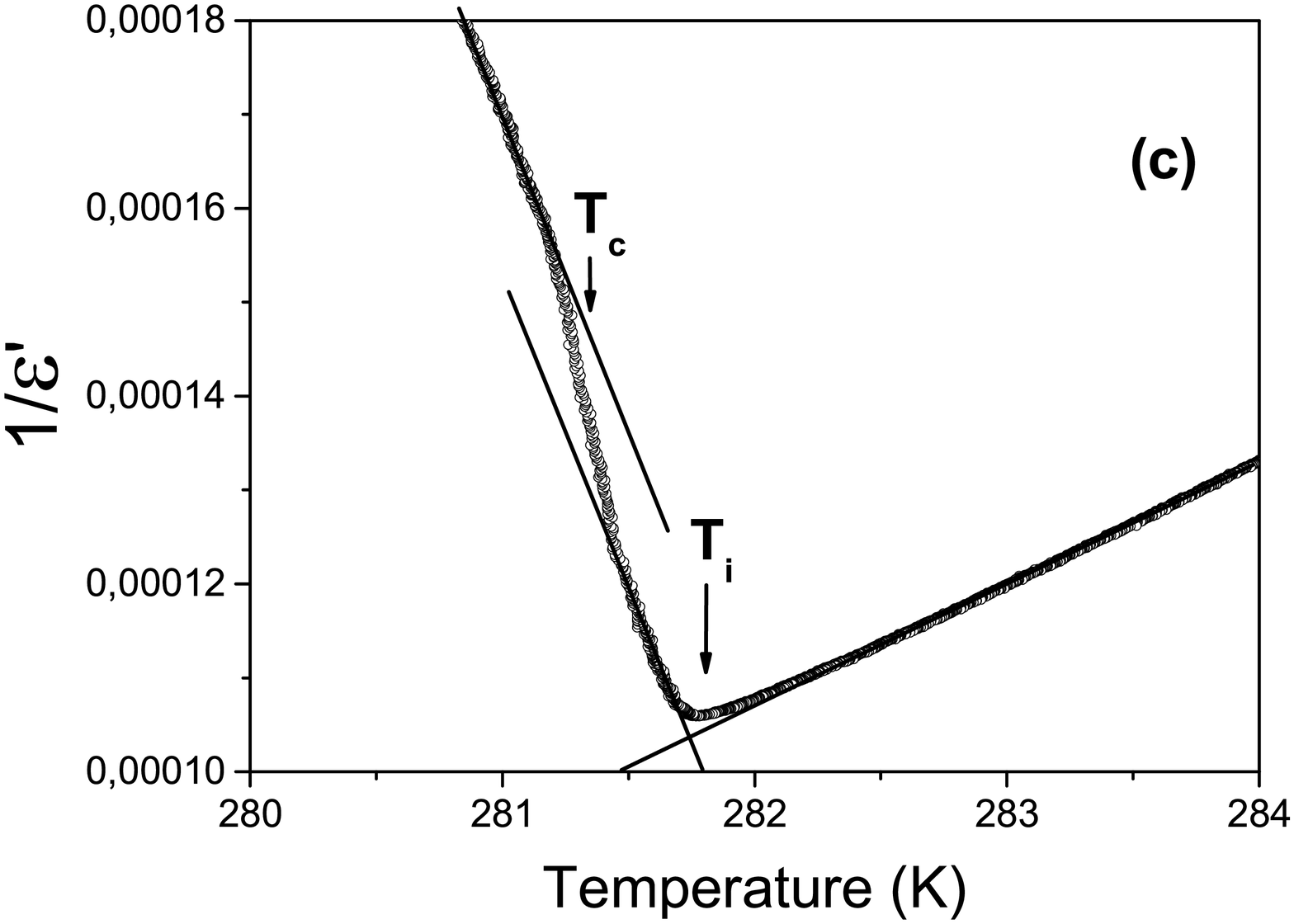}%
\includegraphics[width=8.6cm]{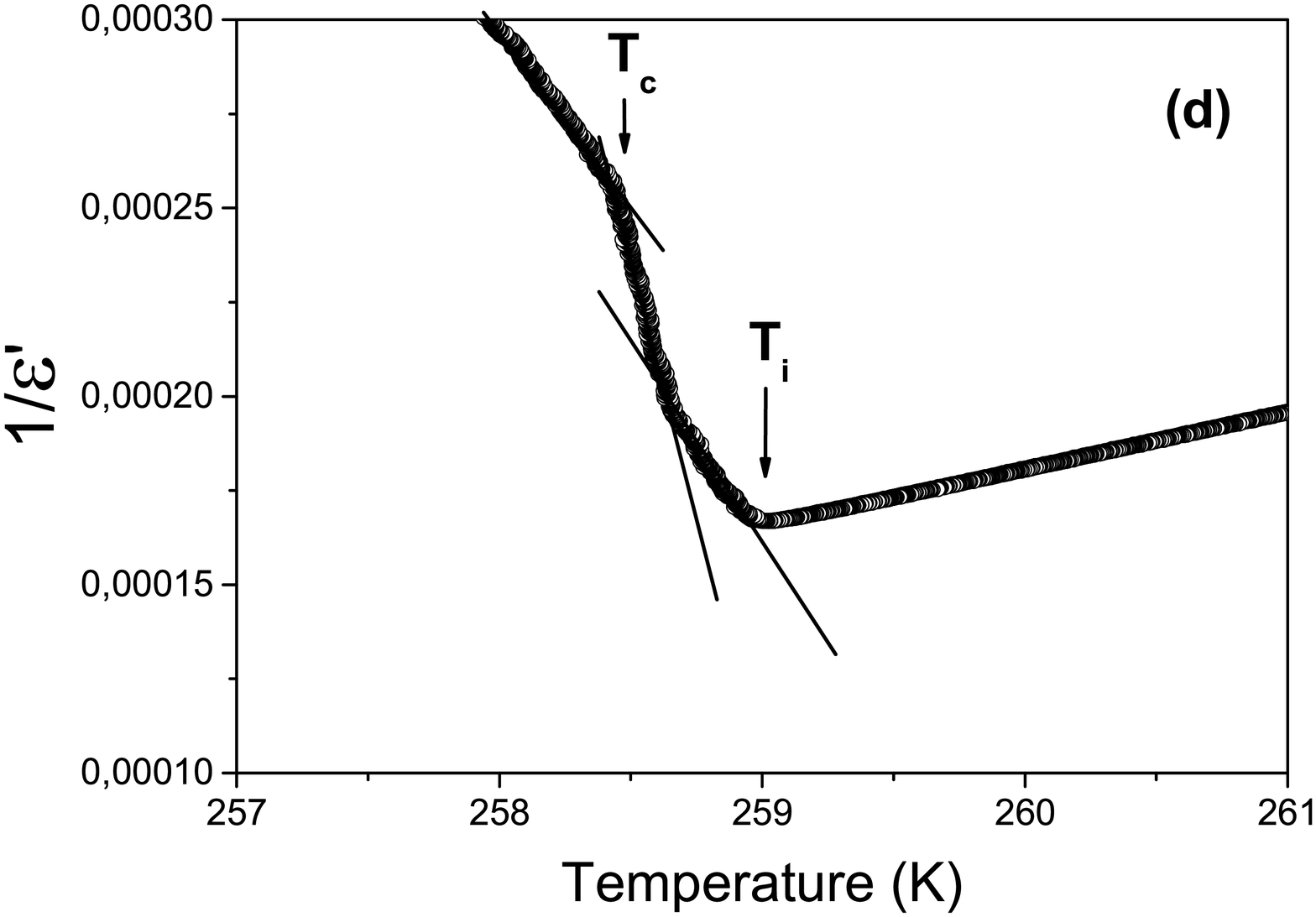} \\
\begin{minipage}[c]{0.5\textwidth}
\includegraphics[width=8.6cm]{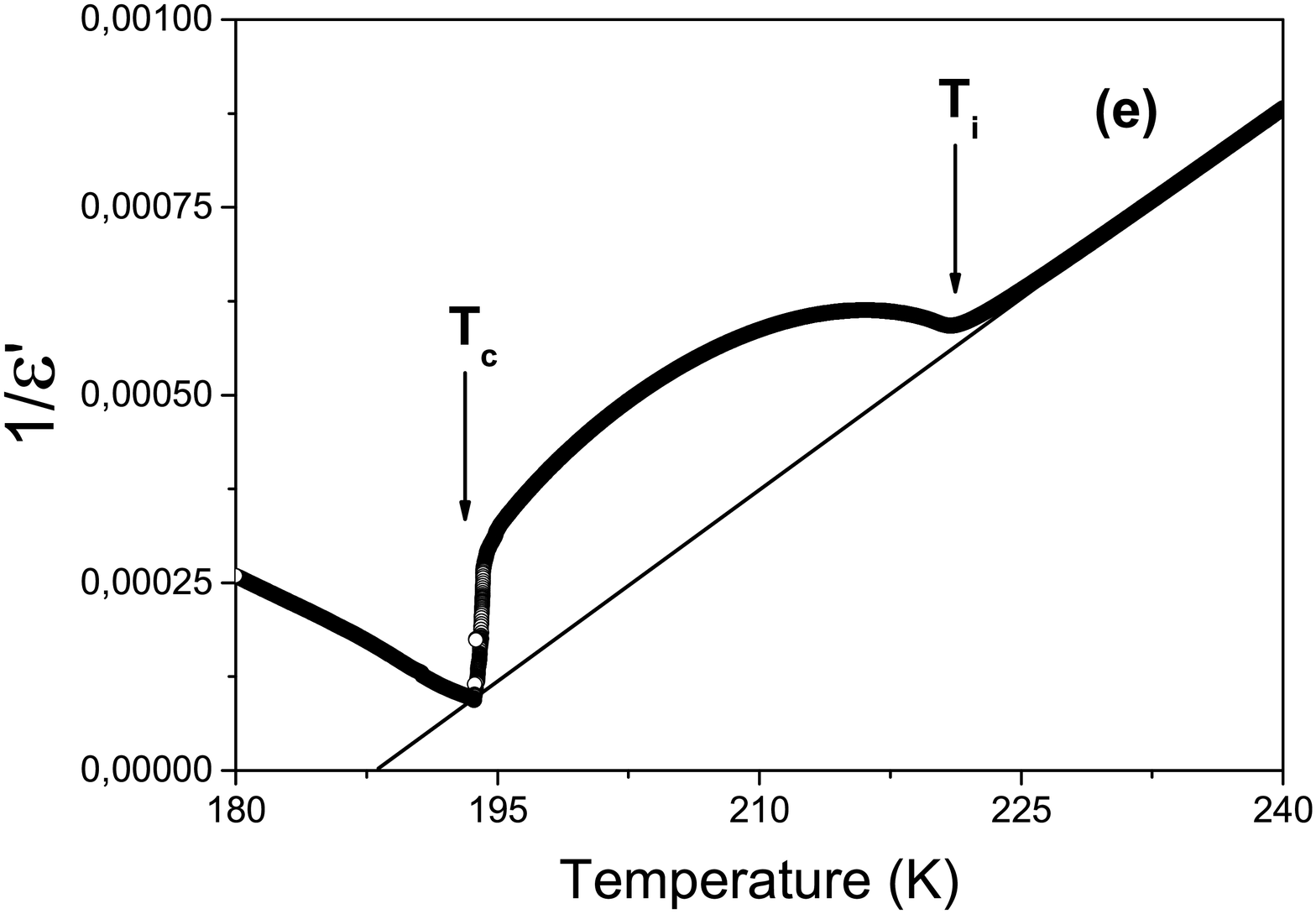} \\
  \end{minipage}\hfill
  \begin{minipage}[c]{0.48\textwidth}
    \caption{Temperature dependence of reciprocal dielectric susceptibility at 0.005~K/min cooling rate for \SPSSe\ mixed crystals with (a) $x = 0$, (b) $x= 0.22$, (c) $x= 0.28$, (d) $x=0.4$, (e) $x=1$.} \label{Fig10}
  \end{minipage}
\end{figure*}

\begin{figure}
\includegraphics[width=8.6cm]{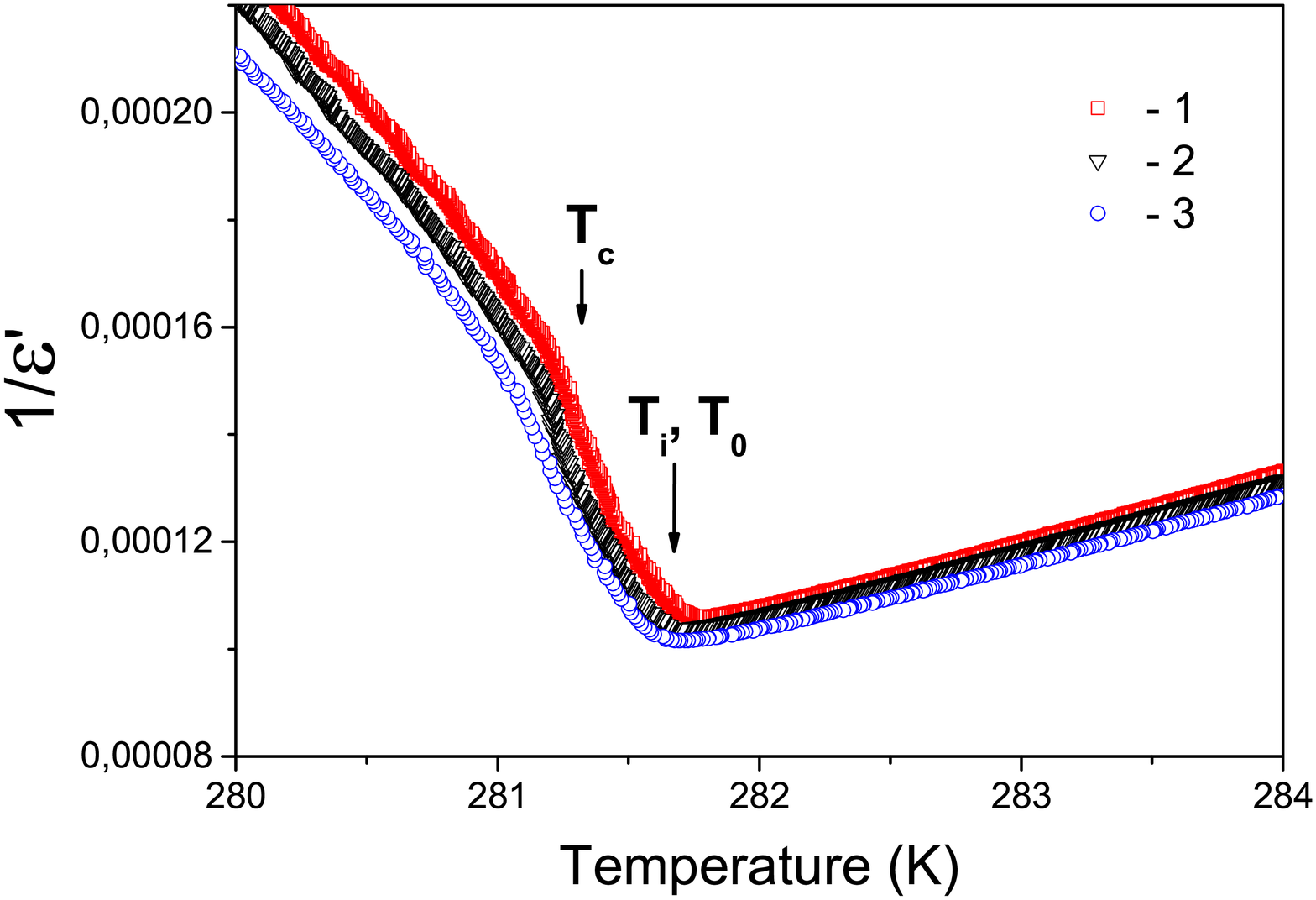} \\
\caption{Temperature dependence of reciprocal dielectric susceptibility for \SPSSe\ mixed crystal with $x = 0.28$ at different cooling rates: (1) 0.005 K/min; (2) 0.05 K/min;  and (3) 0.1 K/min.}
\label{Fig11}
\end{figure}

Recent heat diffusion investigations\cite{ref20} point out, that the Lifshitz point in \SPSSe\ mixed crystals is located near $x = 0.28$. The critical exponents and ratio of the critical amplitudes of heat capacity in paraelectric and ferroelctric phases satisfy predicted critical behavior at the LP in systems without long range interactions.\cite{ref20} Such peculiarity was related to possible screening of dipole-dipole interactions in semiconducting \SPSSe\  with small band gap.\cite{ref20} True critical behavior is observed in very narrow  ($8 \cdot 10^{-3}$ to  $5 \cdot 10^{-5}$) range of reduced temperature $(T -T_0)/T_0$, or in temperature interval $T - T_0$ from 2.4~K to 0.015~K relative to the transition temperature $T_0$. In case of \SPS\  crystals the critical behavior for birefringence, \cite{ref39} dielectric susceptibility,\cite{ref40} ultrasound velocity, \cite{ref41} and heat capacity on thermal diffusion data\cite{ref42} satisfies the mean field exponents with small multiplicative logarithmic corrections. \cite{ref19} Therefore, we could expect, that for analysis of the $T - x$ diagram in \SPSSe\  mixed crystals the mean field values $\Phi  = 0.5$ and $\beta_q = 0.5$ should be appropriate.

According to the experimental data (see Figure~3) for selenium compound ($x = 1$)the temperature difference $T_i - T_c$ is $ \sim 28$~K. As it was discussed above, the LP has position at $x = 0.28$ or at a bit smaller selenium concentration.

According to the relation $T_i - T_c \sim (x - x_{\rm LP})^2$ one could be estimated, that for the mixed crystal with chemical content $x = 0.4$ the temperature width of IC phase $T_i - T_c$ is equal to 0.78~K. In case we assume that the LP is located at  $x_{\rm LP} = 0.26$ similar  estimation gives the temperature interval $T_i - T_c$  equal to 1.00~K for the same concentration $x = 0.4$. Therefore, for sample with $x = 0.28$ the IC phase is predicted to be observed in very small temperature range  $T_i - T_c = 0.02$~K.

From the temperature anomalies in dielectric susceptibility (see Figure~10) it follows, that temperature interval of the IC phase is about 0.6 K for $x = 0.4$, and is about 0.2 K for $x = 0.28$. Experimentally observed temperature range of the IC phase for $x = 0.4$ is satisfied whenwe assume the LP coordinate to be $x_{\rm LP} = 0.28$. However, much wider experimentally observed temperature range (about 0.2 K) of IC phase for $x = 0.28$ sample doesn't agree with such estimations.  The estimated interval $T_i - T_c$ reaches the value of 0.17~K at the $x = 0.28$ only when we suppose that the LP is located at $x_{\rm LP} = 0.22$.

As it follows from above described estimations, for the sample with $x = 0.4$ a true IC phase with temperature width near 0.6~K is observed (see Figure~10). In mixed crystal with  $x = 0.28$ it is still observed an intermediate state between paraelectric and ferroelectric phases, which obviously could be related to some space interference of the long-period modulation wave with regular domain structure with very high concentration of domain walls. This space interference is seen in temperature behavior of the reciprocal dielectric susceptibility for the $x = 0.28$ sample at different cooling rates (se Figure~11): For the lowest rate 0.005~K/min the named dependence is similar to observed for the $x = 0.4$ sample. For the highest rate 0.1~K/min this dependence becomes similar to the anomaly shape in case of the $x = 0.22$ sample (compare with Figure~10).

As it was mentioned above, \SPS\ and \SPSe\ crystals are proper uniaxial ferroelectrics. \SPSe\ has IC phase with almost transverse long-wave modulation, as a result of specific interatomic interaction. This IC phase is not related to the symmetry reason, therefore  Lifshitz invariant is absent in thermodynamic functional. Here, the modulation in the IC phase of Type II is almost harmonic. \cite{ref43} At least, any evidences of higher harmonics were found in the  x-ray and neutron diffraction experiments. \cite{ref13,ref14} The first-order lock-in transition between IC and ferroelectric phase was clearly observed. \cite{ref20,ref37}

At low temperature edge of the IC phase a temperature behavior of some thermodynamic properties (namely, heat capacity, dielectric susceptibility, heat expansion) deviates  from the ones predicted in one-harmonic approximation.  The discrepancies still exist even when the higher harmonics in the spatial modulation of the order parameter in the IC phase are taken into account. This was observed as quantitative difference of  calculated and experimental temperature behavior of thermodynamic properties and modulation wavenumber  in the IC near the lock-in transition. \cite{ref45} Only recently these experimental data  have been successfully explained when higher order invariants (including eight and ten powers) as well as biquadratic coupling of the order parameter with its space derivative were accounted in Landau functional. \cite{ref7,ref8} Such explanation agrees well with strongly anharmonic three-well local potential of \SPS-like ferroelectrics, which was obtained by \textit{ab inito} study of their electronic and dynamical properties. \cite{ref6}

Strong nonlinearity of SPS ferroelectrics allows possibility of specific domain structure with wide domain walls, which include non-polar regions.\cite{ref21}
In order to simulate the influence of the cooling rate on the configuration of the domains in \SPS\ we performed the
Monte Carlo studies of the effective three-well potential Hamiltonian obtained from the first-principles investigations.\cite{ref6}
This Hamlitonian depends on the amplitude of the local mode, which in our case describes atomic displacements of two low-energy optical modes: a polar B$_{\rm u}$ mode, for which the Sn cations are moved out-of-phase to the anionic [P$_2$S$_6$]$^{2-}$ complexes, and full-symmetry A$_{\rm g}$ mode, which describes out-of-phase displacements in the Sn sublattices. The effective Hamiltonian explicitly accounts for the following interactions of the local modes: (i) a self-interaction, which in our case is strongly anharmonic (see discussion in Ref.~\onlinecite{ref6}); (ii) short-range inter-cell interaction; (iii) long-range Coulomb interaction; (iv) elastic energy; and (v) anisotropic coupling of the local mode to elastic deformations.
In Monte Carlo simulations we used $1 \times 200 \times 1$ supercell. This means, that we allowed evolution of the domains in crystallographic direction $b$ (as it is observed experimentally), whereas in other directions we assumed mean-field behavior.
This effective Hamiltonian was solved in cooling regime starting from 600~K with temperature step of 5~K down to 10~K.
Variations of the cooling rate were simulated by different numbers of pseudospin updates. In so called \textit{fast} cooling we used 10$^5$ updates for each pseudospin both for equilibration and production cycles. \textit{Slow} cooling was simulated by 10$^6$ updates.
The temperature evolution of domain structure at different cooling rates and several temperatures is shown at Figure~12.
It is clearly seen, that in the vicinity of the phase transition temperature ($\sim$337~K) domain configuration obtained in the \textit{fast} regime reveals short-period microdomains. This is in contrast to the \textit{slow} cooling, where resulted domain configuration is almost harmonic with the period of the simulation cell. The effect of the cooling rate on domain configuration is important only in the vicinity of the phase transition. By further cooling, the domain structure in both cases reveal two clearly distinguished areas with opposite directions of the polarization.
Sharp domain walls with non-polar areas are observed at very low temperatures.

We should note, that similar exotic temperature evolution of the domain structure with high flexibility was recently predicted by MC simulation and consequently confirmed in piezoresponse force microscopy (PFM) experiments. \cite{ref21}

Increasing the domain walls concentration by heating in ferroelectric phase was found in \SPSe\ crystal analytically and in phase-field modeling. \cite{ref2} It seems, that  below the lock-in transition temperature $T_c$ the mean size of domains is small enough and domain walls width is very big, i.e. domain structure becomes similar to the periodically modulated space distribution of the spontaneous polarization in the IC phase. Possible similarity of the  space distribution of spontaneous polarization just below $T_c$ and above this temperature also is supported by the presence of small kink in the optic birefringence at lock-in transition in \SPSe. \cite{ref46} High dielectric response of the domain walls below lock-in transition (See Figure~1) is evidently also related to their high concentration.

  \begin{figure*}
\includegraphics[width=8.6cm]{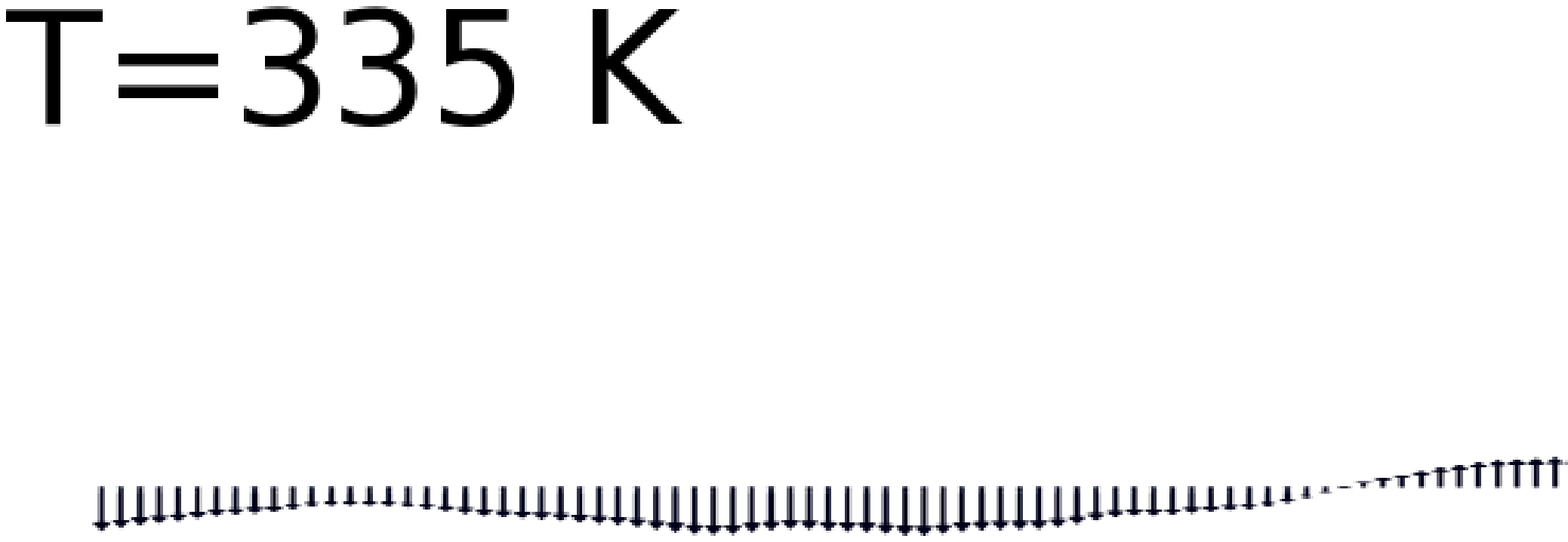}%
\includegraphics[width=8.6cm]{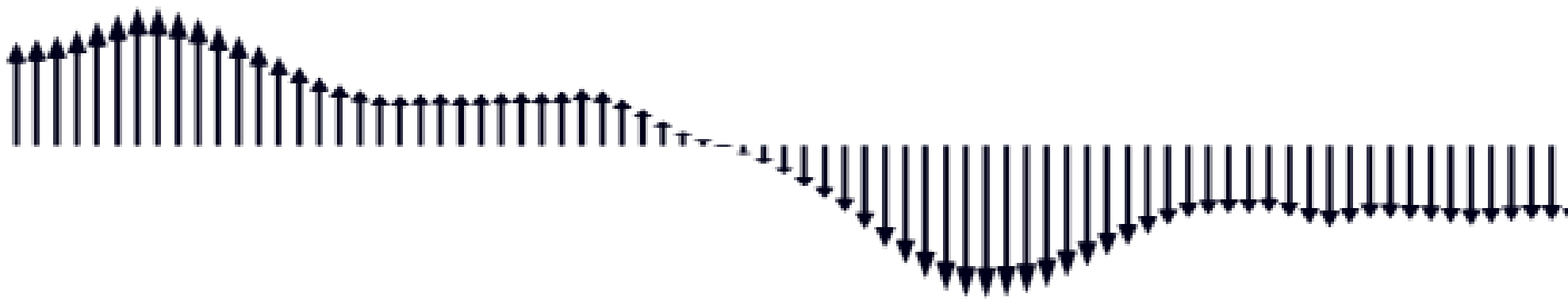} \\
\includegraphics[width=8.6cm]{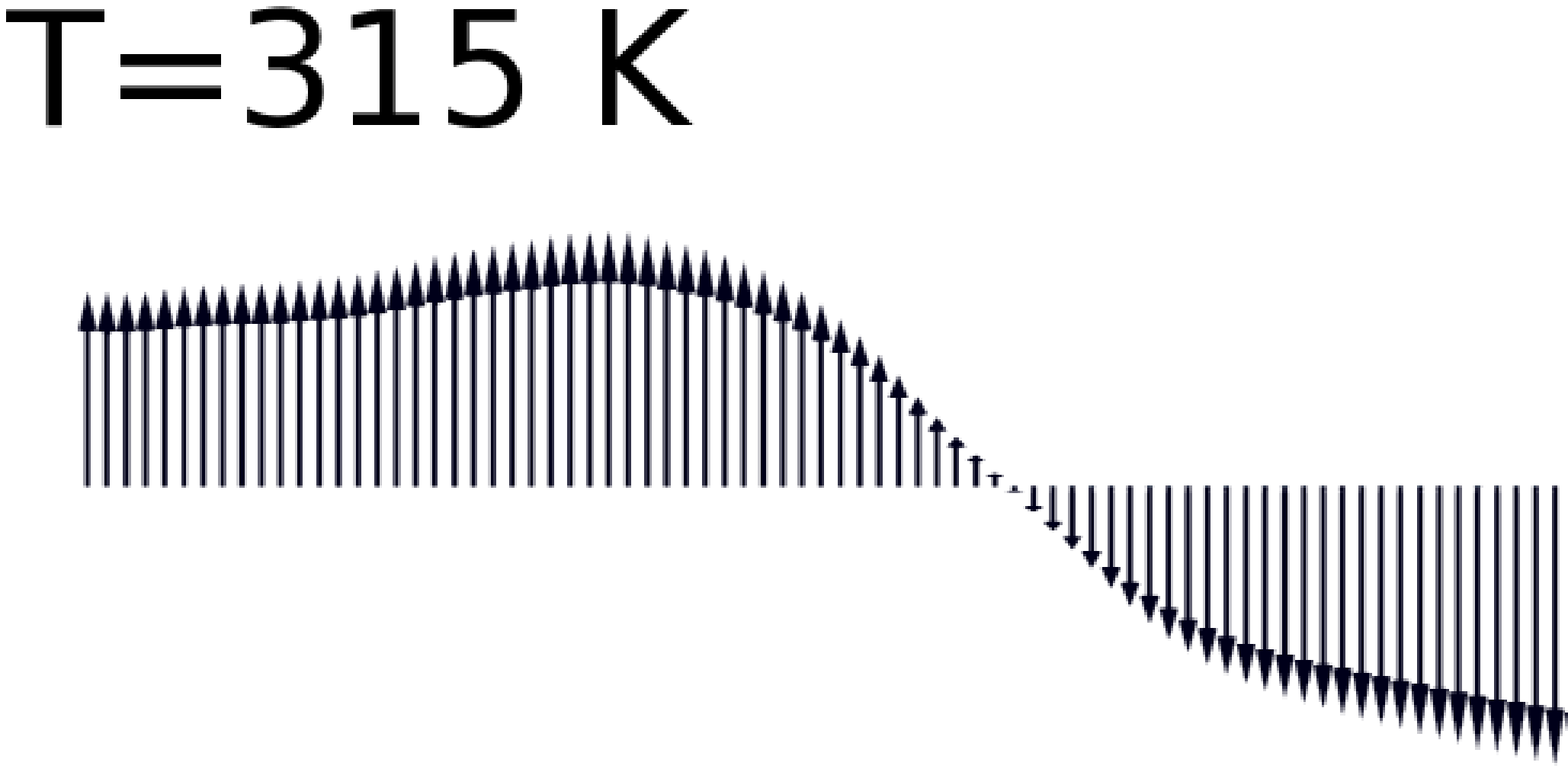}
\includegraphics[width=8.6cm]{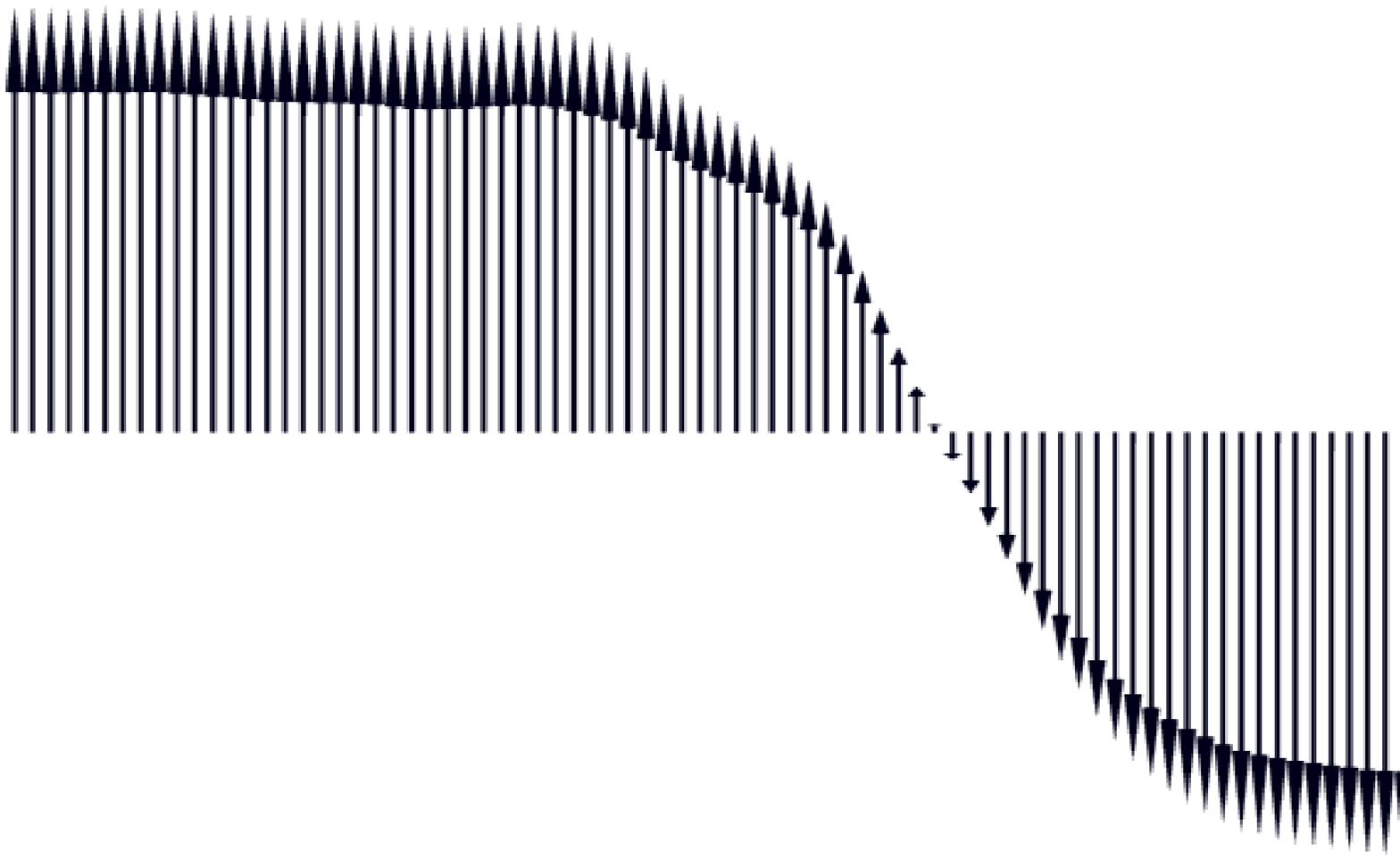} \\
\includegraphics[width=8.6cm]{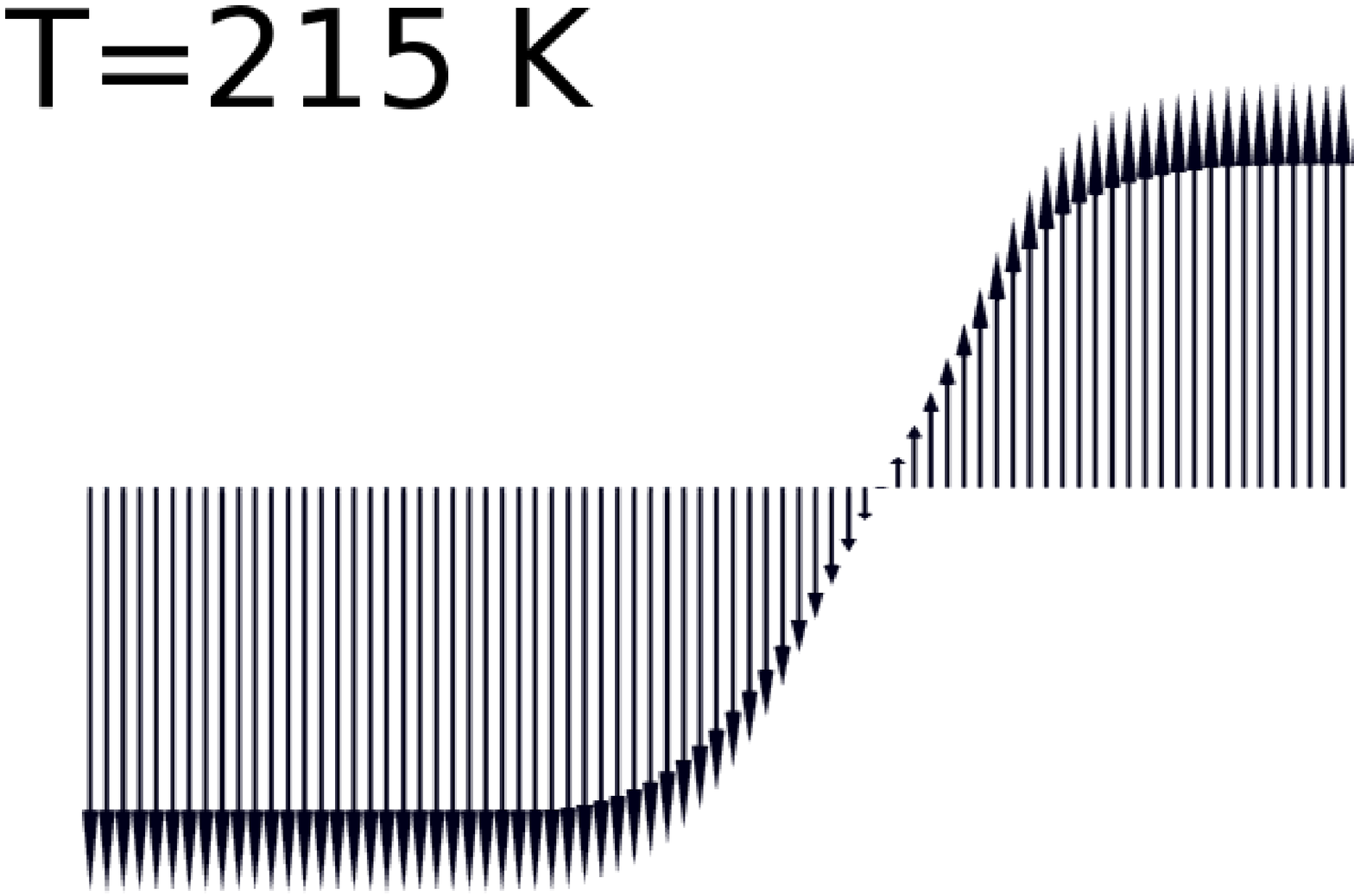}
\includegraphics[width=8.6cm]{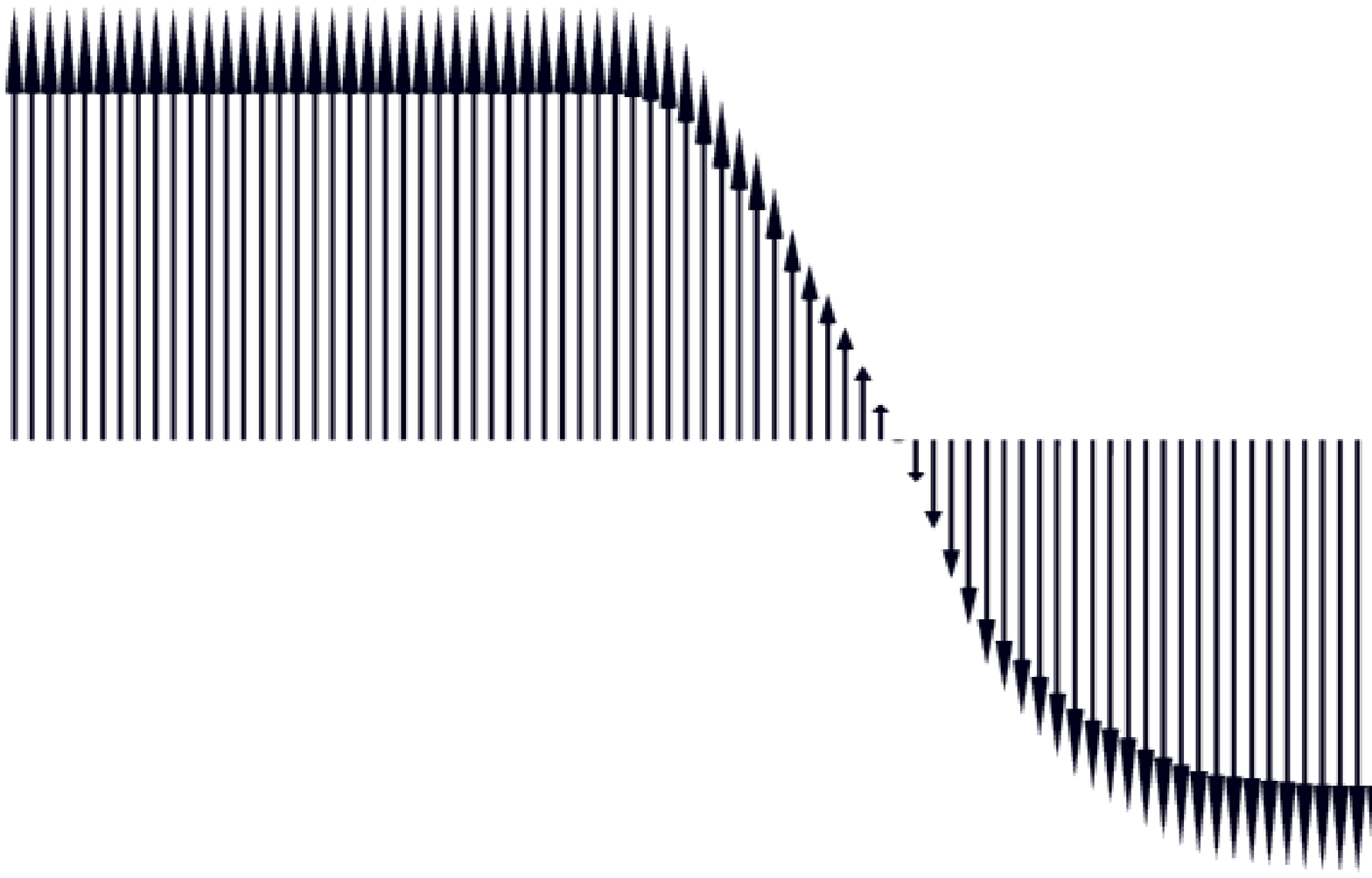} \\
\includegraphics[width=8.6cm]{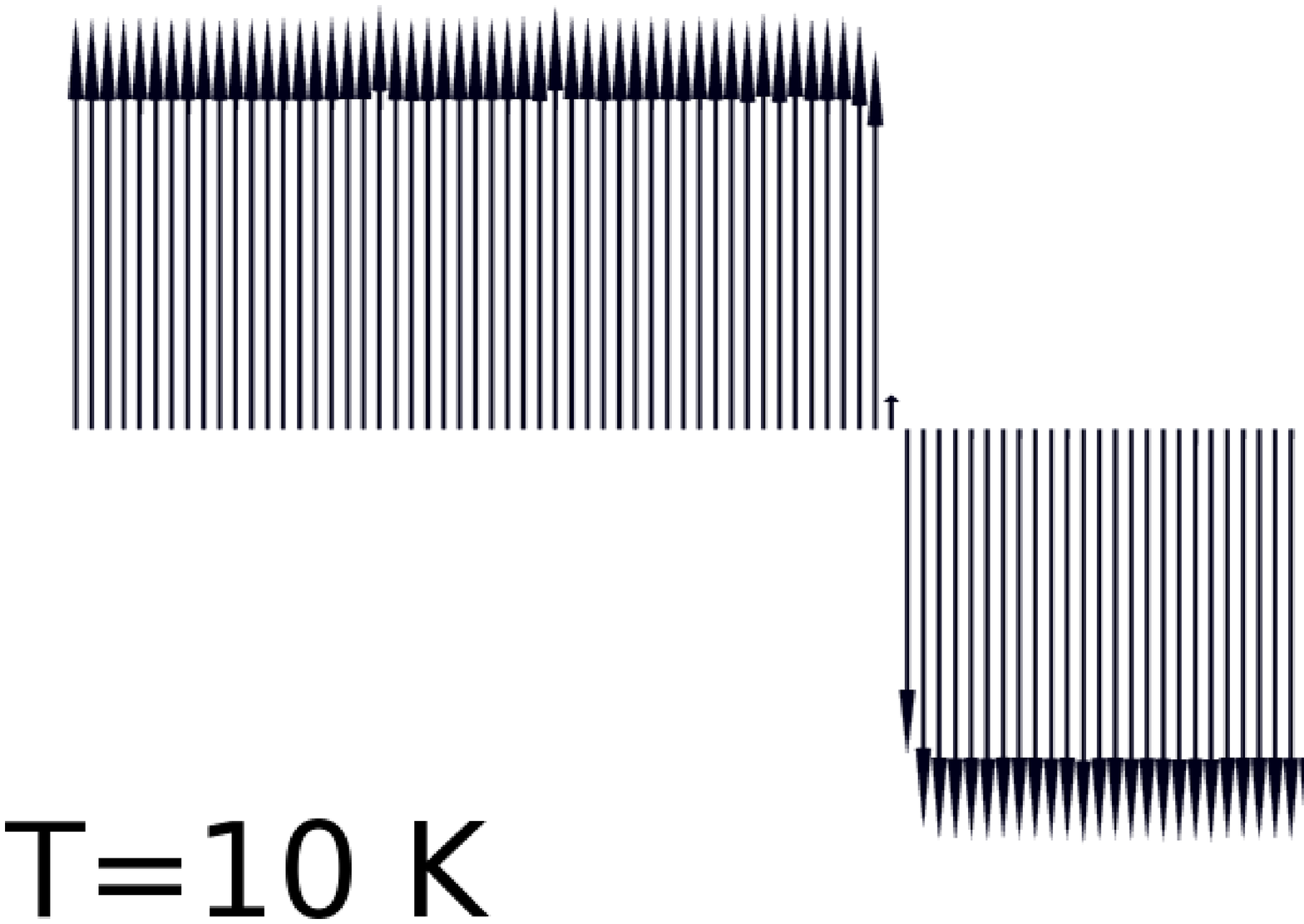}
\includegraphics[width=8.6cm]{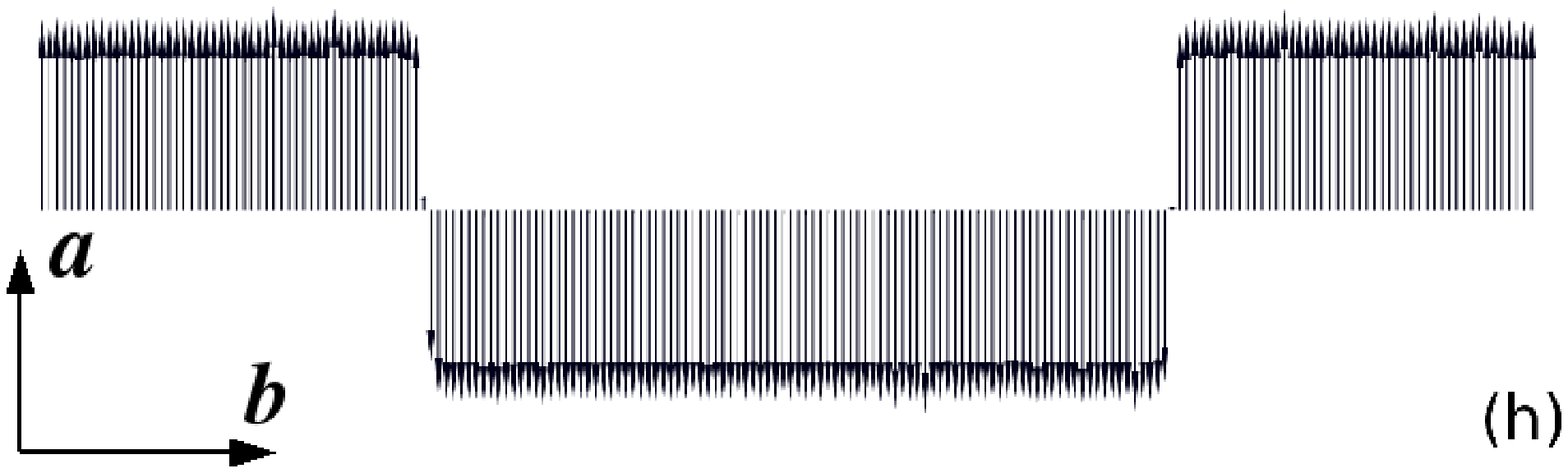} \\
\caption{Monte Carlo domain structure of \SPS\ at various temperatures and different length of Markov chains in cooling regime: (a), (c), (e), (g) corresponds to $10^6$ sweeps per pseudospin (and, correspondingly, \textit{slow} cooling rate), whereas (b), (d), (f), (h) results are obtained for $10^5$ sweeps per pseudospin (i.e., the \textit{fast} cooling). Direction of the polarization is mostly along crystallographic period $a$. Domain structure appears along crystallographic direction $b$. }
\label{Fig12}
\end{figure*}

When the LP is approached along $T_c(x)$ line the modulation period of IC phase growth and at some conditions could become comparable with domains dimensions in ferroelectric phase near the LP. Also, the temperature range $T_i - T_c$ of IC phase disappears when approaching to the LP. Obviously, just above $T_c$ the modulation wave is expected to be harmonic. Some kink in dielectric properties at $T_c$ near the LP is related to small difference in space profile of  spontaneous polarization in IC and ferroelectric phases. The concentration of domain walls  depends on the cooling rate.  Therefore, the observed kink at $T_c$ in the dielectric susceptibility could be changed when cooling rate is varied.

Next, we will try to discuss the domain walls in uniaxial ferroelectrics as scalar topological defects \cite{ref29,ref30,ref31} with concentration, that depends on the cooling rate across second order transition from paraelectric phase to the ferroelectric one. When a system goes through a symmetry breaking phase transition from a symmetric phase into one with spontaneously broken symmetry, the order parameter may make different choices in different regions, creating polar domains. Domains with different direction of polarization, when they meet, can create defects such as the domain walls. The scale of those domains, and hence the density of defects, is constrained by the speed at which the system goes through the transition and the speed with which order parameter information propagates. At slow cooling rate different regions can propagate their choice of phase: large regions found the same choice and low density of defects is presented. At fast cooling rate there is less time to communicate the choice of phases. Therefore, many small regions with different choice of phases appear, i.e.  density of defects is high.

By application for ferroelectrics it was argued, \cite{ref31} that for fast cooling through the phase transition point, the distance over which information can be transferred is short, and becomes equal to the smallest correlation length $\xi(T)$. Therefore, freeze-out occurs when the domain size is small and consequently the concentration of topological defects, which are domain walls $n_w$,  is large. In contrast, for slow cooling, the distance for information transfer is large and does not become equal to $\xi(T)$ until  the phase transition temperature, where $\xi(T)$ is large. In this case, large domains are formed, and a few topological defects like domain walls are observed.
Quantitatively, the Kibble-Zurek theory predicts the following dependence of  domain size $d$ on cooling speed: \cite{ref29,ref30}:
\begin{equation}
d=\xi_0 (\dfrac{\tau_q}{\tau_0})^{\frac{\nu}{1+\mu}},
\end{equation}
 			
\noindent where $\xi_0$ -- zero temperature correlation length, that is proportional to the domain walls width; $\tau_q = \frac{T_0}{\nu_T}$ -- a ratio of the PT temperature $T_0$ to the cooling speed $\nu_T = \frac{dT}{dt}$;  $\tau_0 = \frac{\xi_0}{\nu_s}$ is zero temperature relaxation time, where $\nu_s$ is speed of sound. The critical indexes $\mu$ and $\nu$ determine divergence of correlation length and relaxation time at transition temperature:
\begin{equation}
\xi(T)=\xi_0 (1-\dfrac{\tau}{\tau_0})^{-\nu}; \ \ \ \tau(T)=\tau_0 (1-\dfrac{\tau}{\tau_0})^{-\mu}.
\end{equation}

In the vicinity of the LP in uniaxial ferroelectrics \SPSSe\  near to the tricritical point on $T - x$ diagram, \cite{ref5,ref16} the universality class of the UTLP could be appropriated. Such policritical point is described by mean-field critical indexes with small multiplicative logarithmic corrections\cite{ref19} and values $\nu = 0.5$ and $\mu = 1$ could be used for estimations. As follows from the DFT calculations, the lowest energy domain walls width in  \SPS\ crystals   is near two elementary cells,\cite{ref6} i.e. $\xi_0\approx 20$~\AA. The speed of transverse acoustic waves for the mixed crystals \SPSSe\ with $x = 0.28$ was experimentally determined as $\nu_s \approx 2000$~m/s.  \cite{ref25}

With this set of parameters from relation (1) follows, that at cooling speed $\frac{dT}{dt} = 0.005$ K/min  the domains dimension $d$ is near 230~$\mu m$. When cooling speed is 0.1~K/min, value of $d$ lowers to $ \approx 50$ $\mu m$ that is comparable with several micrometers modulation wave length $\lambda_i = \dfrac{2\pi}{q_i}$ for $q_i \approx10^{-3}$ \AA$^{-1}$ near the LP at concentration distance $x - x_{\rm LP} \approx 0.01$ (see Figure~4).

\section{CONCLUSIONS}

For mixed crystals \SPSSe\ with selenium concentration near the Lifshitz point $x_{\rm LP} \sim 0.28$ the anomalies of dielectric susceptibility demonstrate different temperature width of the incommensurate phase,  depending on the cooling rates. Observed  dependence of lock-in transition on the cooling rate could be associated to transformation of the long-wave modulation of polarization into domain structure with wide enough domain walls, which are determined by strongly nonlinear local potential and have different concentration according to the the cooling rate. For composition $x = 0.28$ the intermediate IC phase, with temperature interval between $T_i$ and $T_c$ about 0.1~K, is observed in the regime of the slowest cooling speed 0.002~K/min. The kink in dielectric susceptibility at lock-in transition $T_c$ smears by increasing the cooling rate to 0.1~K/min. This smearing is related to increase of the domain wall concentration $n_w$ in the ferroelectric phase just below $T_c$. Such effect of nonequilibrium behavior near the Lifshitz point is described within known Kibble-Zurek model.\cite{ref29,ref30} For the uniaxial ferroelectrics the domain walls could be considered as scalar topological defects with concentration $n_w$, which depends on the cooling rate. We suggest, that according to KZ relations,\cite{ref29,ref30}  $n_w$ strongly growth when cooling rates increase from 0.002~K/min to 0.1~K/min. Biggest value $n_w$ gives distance between the domain walls near 50~nm. This value is comparable with modulation wave length, which is expected near the LP at concentration distance $x - x_{\rm LP} \sim 0.01$.

\begin{acknowledgments}
K.Z.R. is grateful to the Alexander von Humboldt Foundation and Helmholtz Young Investigators Group Programme VH-NG-409 for financial support.
\end{acknowledgments}

\bibliography{SPS_KZ}

\end{document}